\documentclass[
superscriptaddress,
final,
reprint,
showkeys,
pre,
aps,
floatfix,
]{revtex4-1}

\usepackage{nameref}
\usepackage[dvips]{graphicx}
\usepackage[utf8]{inputenc}
\usepackage{multirow}
\usepackage{amsmath,amssymb,latexsym,MnSymbol,bm}

\usepackage[table]{xcolor}
\usepackage{xcolor}
\usepackage[%
  colorlinks=true,
  urlcolor=blue,
  linkcolor=blue,
  citecolor=blue
]{hyperref}
\usepackage[final]{pdfpages}
\makeatletter
\AtBeginDocument{\let\LS@rot\@undefined}
\makeatother

\begin{document}

\title{City size and the spreading of COVID-19 in Brazil}

\author{Haroldo\ V.\ Ribeiro}
\email{hvr@dfi.uem.br}
\affiliation{Departamento de F\'isica, Universidade Estadual de Maring\'a -- Maring\'a, PR 87020-900, Brazil}

\author{Andre\ S.\ Sunahara}
\affiliation{Departamento de F\'isica, Universidade Estadual de Maring\'a -- Maring\'a, PR 87020-900, Brazil}

\author{Jack~Sutton}
\affiliation{School of Science and Technology, Nottingham Trent University, Clifton Lane, Nottingham NG11 8NS, United Kingdom}

\author{Matja{\v z}\ Perc}
\affiliation{Faculty of Natural Sciences and Mathematics, University of Maribor, Koro{\v s}ka cesta 160, 2000 Maribor, Slovenia}
\affiliation{Department of Medical Research, China Medical University Hospital, China Medical University, Taichung, Taiwan}
\affiliation{Complexity Science Hub Vienna, Josefst{\"a}dterstra{\ss}e 39, 1080 Vienna, Austria}

\author{Quentin\ S.\ Hanley}
\affiliation{School of Science and Technology, Nottingham Trent University, Clifton Lane, Nottingham NG11 8NS, United Kingdom}

\date{\today}
\begin{abstract}
The current outbreak of the coronavirus disease 2019 (COVID-19) is an unprecedented example of how fast an infectious disease can spread around the globe (especially in urban areas) and the enormous impact it causes on public health and socio-economic activities. Despite the recent surge of investigations about different aspects of the COVID-19 pandemic, we still know little about the effects of city size on the propagation of this disease in urban areas. Here we investigate how the number of cases and deaths by COVID-19 scale with the population of Brazilian cities. Our results indicate small towns are proportionally more affected by COVID-19 during the initial spread of the disease, such that the cumulative numbers of cases and deaths \textit{per capita} initially decrease with population size. However, during the long-term course of the pandemic, this urban advantage vanishes and large cities start to exhibit higher incidence of cases and deaths, such that every $1\%$ rise in population is associated with a $0.14\%$ increase in the number of fatalities \textit{per capita} after about four months since the first two daily deaths. We argue that these patterns may be related to the existence of proportionally more health infrastructure in the largest cities and a lower proportion of older adults in large urban areas. We also find the initial growth rate of cases and deaths to be higher in large cities; however, these growth rates tend to decrease in large cities and to increase in small ones over time.
\end{abstract}

\maketitle

\section*{Introduction}
Human activities have become increasingly concentrated in urban areas. A direct consequence of this worldwide urbanization process is that more people are living in cities than in rural regions since 2007~\cite{WorldBankPop}, and projections indicate that the world urban population could reach more than 90\% by the end of this century~\cite{jiangglobal2017}. Besides being increasingly urbanized, we live in an unprecedentedly connected, and highly mobile world with air passengers exceeding 4 billion in 2018~\cite{WorldBankAir}. On the one hand, a highly connected and highly urbanized society brought us innovation, economic growth, more access to education and healthcare; on the other, it has also lead to pollution, environmental degradation, privacy concerns, more people living in substandard conditions, and suitable conditions for dissemination of infectious diseases over the globe. In particular, the emergence of infectious disease outbreaks has significantly increased over time, and the majority of these events are caused by pathogens originating in wildlife~\cite{jones2008global}, which in turn has been associated with changes in environmental conditions and land use, agricultural practices, and the rise of large human population settlements~\cite{wolfe2007origins}.

The ongoing outbreak of the novel coronavirus (SARS-CoV-2) seems to fit well the previous context as it was first identified in Wuhan in December 2019, an influential Chinese city exceeding 11 million inhabitants, and apparently originated from the recombination of bat and Malayan pangolin coronaviruses~\cite{xiao2020isolation}. The coronavirus disease 2019 (COVID-19) initially spread in Mainland China but rapidly caused outbreaks in other countries, making the World Health Organization first declare a ``Public Health Emergency of International Concern'' in January 2020, and in mid-March, the outbreak was reclassified as a pandemic. As of 16 August 2020, over 21.2 million cases of COVID-19 have been confirmed in almost all countries, and the worldwide death toll exceeds 761 thousand people~\cite{WHO_COVID19}. The COVID-19 pandemic poses unprecedented health and economic threats to our society, and understanding its spreading patterns may find important factors for mitigating or controlling the outbreak. 

Recent works have focused on modeling the initial spreading of COVID-19~\cite{maier2020effective} or the fatality curves~\cite{vasconcelos2020modelling}, projecting the outbreak peak and hospital utilization~\cite{moghadas2020projecting}, understanding the effects of mobility~\cite{gatto2020spread}, demography~\cite{dowd2020demographic}, travel restrictions~\cite{chinazzi2020effect}, behavior change on the virus transmission~\cite{west2020applying}, mitigation strategies~\cite{walker2020impact}, non-pharmaceutical interventions~\cite{flaxman2020estimating}, network-based strategies for social distancing~\cite{block2020social}, among many others. Despite the increasing surge of scientific investigations on the subject, little attention has been paid to understanding the effects of city size on spreading patterns of cases and deaths by COVID-19 in urban areas. The idea that size (as measured by population) affects different city indicators has been extensively studied and can be summarized by the urban scaling hypothesis~\cite{bettencourtgrowth2007,bettencourt2013origins,battynew2013,westscale2017}. This theory states that urban indicators are non-linearly associated with city population such that socio-economic indicators tend to present increasing returns to scale~\cite{bettencourtgrowth2007,youn2016scaling,gao2019computational}, infrastructure indicators often display economy of scale~\cite{bettencourtgrowth2007,bettencourt2013origins}, and quantities related to individual needs usually scale linearly with city population~\cite{bettencourtgrowth2007,bettencourt2013origins}. 

Urban scaling studies of health-related quantities have shown that the incidence and mortality of diseases are non-linearly related to the city population~\cite{acuna2011influenza,antonio2014growth,melo2014statistical,rocha2015non}. Despite the existence of several exceptions~\cite{rocha2015non}, noninfectious diseases (such as diabetes) are usually less prevalent in large cities, while infectious diseases (such as AIDS) are relatively more common in large urban areas. This different behavior is likely to reflect the fact that people living in large cities tend to have proportionally more contacts and a higher degree of social interactions than those living in small towns~\cite{bettencourt2013origins,schlapfer2014scaling}. Within this context, the recent work of Stier, Berman, and Bettencourt~\cite{stier2020covid} has indicated that large cities in the United States experienced more pronounced growth rates of COVID-19 cases during the first weeks after the introduction of the disease. Similarly, Cardoso and Gon\c{c}alves~\cite{cardoso2020urban} found that the \textit{per capita} contact rate of COVID-19 increases with the size and density of cities in United States, Brazil and Germany. These findings have serious consequences for the evolution of COVID-19 and suggest that large metropolises may become infection hubs with potentially higher and earlier peaks of infected people. Investigating whether this behavior generalizes to other places and how different quantities such as the number of cases and deaths scale with city size are thus important elements for a better understanding of the spreading of COVID-19 in urban areas.

Here we investigate how population size is associated with cases and deaths by COVID-19 in Brazilian cities. Brazil is the sixth most populous country in the world, with over 211 million people, of which more than 85\% live in urban areas. While it is likely that the novel coronavirus was already circulating in Brazil in early February 2020~\cite{delatorre2020tracking}, the first confirmed case in the country dates back to 26 February 2020, in the city of S\~ao Paulo. Between the first case and 12 August 2020, Brazil has confirmed 3,088,670 cases of COVID-19 (second-largest number) spread out over 98.9\% of the 5,570 Brazilian cities. This disease caused 102,817 deaths (second-largest number) with 3,892 cities reporting at least one casualty as of 12 August 2020.

\section*{Results}

We start by briefly presenting our data set (see Methods for details). Our investigations rely on the daily reports published by the Health Offices of each of the 27 Brazilian federative units. These daily reports update the number of confirmed cases ($Y_c$) and the number of deaths ($Y_d$) caused by COVID-19 in every Brazilian city from 25 February 2020 (date of the first case in Brazil) to 12 August 2020 (date of our last update). From these data, we create time series of the number of cases $Y_c(t_c)$ for each city, where $t_c$ refers to the number of days since the first two daily cases reported in each city. Similarly, we create time series of the number of deaths $Y_d(t_d)$, where $t_d$ refers to the number of days since the first two daily deaths reported in each city. By doing so, we group all cities according to their stage of disease propagation (as measured by $t_c$ or $t_d$) to investigate the evolution of allometric relationships between total cases or deaths and city population. We have also considered different number of daily cases or deaths as the reference point, and our results are robust against different choices (from one to seven daily cases or daily deaths, see Figures~1-14 in S1~Appendix).

\begin{figure*}
    \centering
    \includegraphics[width=1\textwidth]{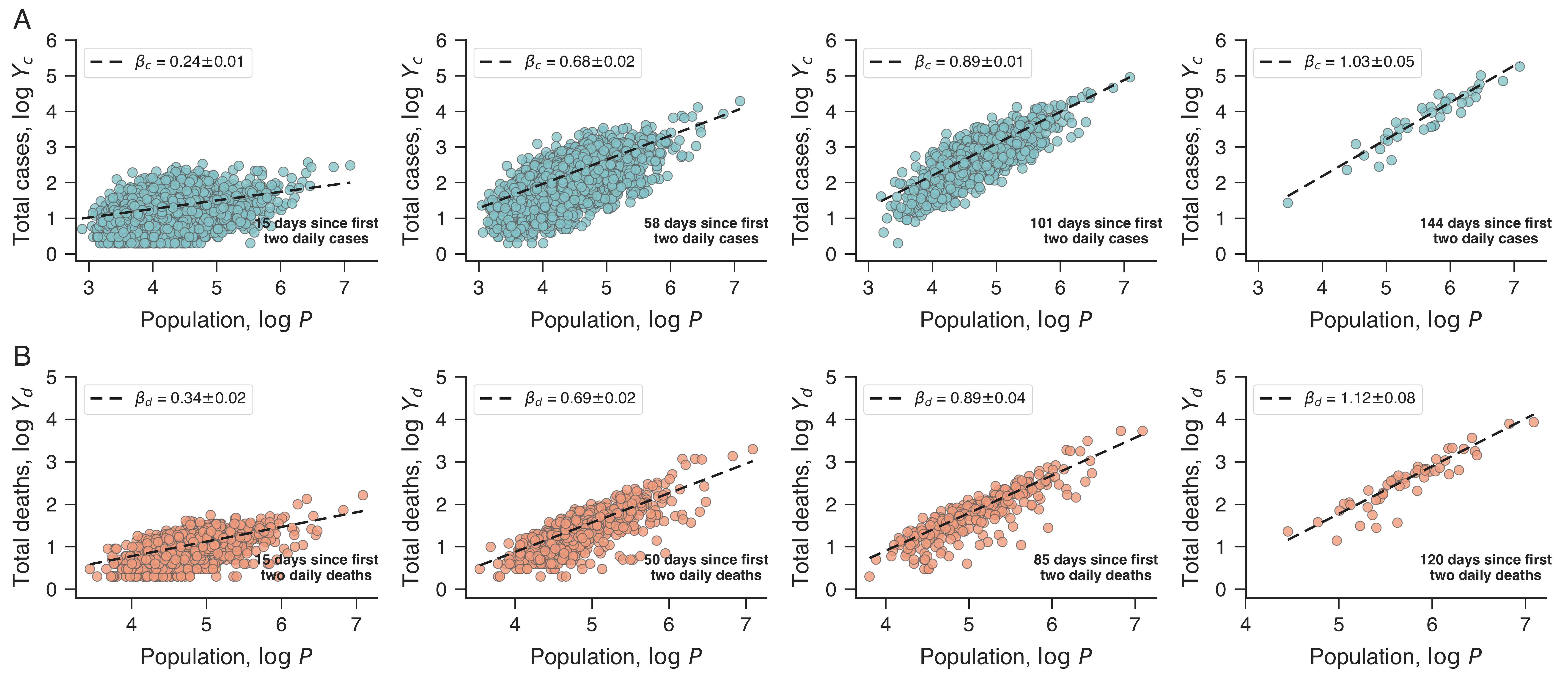}
    \caption{\textbf{Urban scaling relations of COVID-19 cases and deaths.} (A) Relationship between the total of confirmed cases of COVID-19 ($Y_c$) and city population ($P$) on logarithmic scale. Panels show scaling relations for the number of cases on a particular day after the first two daily cases reported in each city (four evenly spaced values of $t_c$ between 15 days and the largest value yielding at least 50 cities, as indicated within panels). (B) Relationship between the total of deaths caused by COVID-19 ($Y_d$) and population ($P$) of Brazilian cities (on logarithmic scale). Panels show scaling relations for the number of deaths on a particular day after the first two daily deaths reported in each city (four evenly spaced values of $t_d$ between 15 days and the largest value yielding at least 50 cities, as indicated within panels). In all panels, the markers represent cities and the dashed lines are the adjusted scaling relations with best-fitting exponents indicated in each plot ($\beta_c$ for cases and $\beta_d$ for deaths).}
    \label{fig:1}
\end{figure*}

Figure~\ref{fig:1}A shows the relation between cases of COVID-19 and city population on a logarithmic scale ($\log Y_c$ versus $\log P$) for different numbers of days since the first two daily cases ($t_c=15,58,101$ and $141$ days). The approximately linear behavior on logarithmic scale indicates that the number of cases is well described by a power-law function of the city population
\begin{equation}\label{eq:scaling_cases}
    Y_c\sim P^{\beta_c}\,,
\end{equation}
where $\beta_c$ is the so-called urban scaling exponent~\cite{bettencourtgrowth2007}. Similarly, Figure~\ref{fig:1}B shows the association between the number of casualties and the city population on logarithmic scale ($\log Y_d$ versus $\log P$) for different numbers of days since two daily deaths first reported ($t_d=15,50,85$ and $120$ days). Again, the results indicate that the number of deaths is approximated by a power-function of the city population
\begin{equation}\label{eq:scaling_deaths}
    Y_d\sim P^{\beta_d}\,,
\end{equation}
where $\beta_d$ represents the urban scaling exponent for the number of deaths. 

\begin{figure*}
    \centering
    \includegraphics[width=0.7\textwidth]{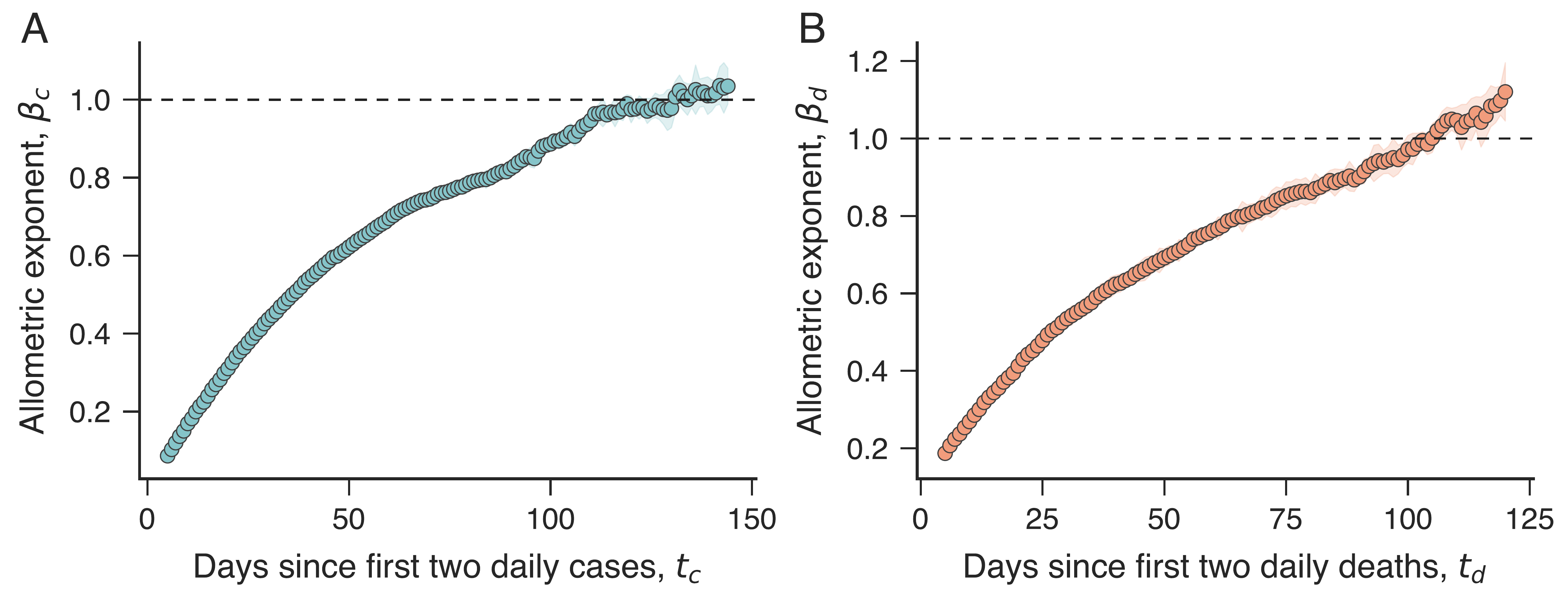}
    \caption{\textbf{Time dependence of the scaling exponents for COVID-19 cases and deaths.} (A)~Dependence of the exponent $\beta_c$ on the number of days after the first two daily cases of COVID-19 ($t_c$). (B) Dependence of the exponent $\beta_d$ on the number of days after the first two daily deaths caused by COVID-19 ($t_d$). The shaded regions in all panels represent bootstrap standard errors, and the horizontal dashed lines indicate the isometric scaling ($\beta_c=\beta_d=1$). We note that $\beta_c$ and $\beta_d$ increase with time and appear to approach asymptotic values larger than one.}
    \label{fig:2}
\end{figure*}

The results of Figure~\ref{fig:1} also show the adjusted allometric relationships (dashed lines) and the best fitting scaling exponents $\beta_c$ and $\beta_d$ (see Methods for datails). These exponents exhibit an increasing trend with time so that $\beta_c$ and $\beta_d$ exceed one after some number of days after the first two daily cases or deaths. This dynamic behavior is better visualized in Figure~\ref{fig:2}, where we depict $\beta_c$ and $\beta_d$ as a function of the number of days since the first two daily cases ($t_c$) or deaths ($t_d$). The scaling exponent for the number of cases $\beta_c$ is sub-linear ($\beta_c<1$) during the first four months and appears to approach a super-linear plateau ($\beta_c>1$) as the number of days $t_c$ further increases. The dynamic behavior of the scaling exponent for deaths $\beta_d$ is similar to $\beta_c$; however, $\beta_d$ appears to be approaching a plateau larger than the one observed for $\beta_c$.

The evolution of the scaling exponents for cases and deaths indicates that small cities are proportionally more affected by COVID-19 during the first four months. However, this initial apparent advantage of living in large cities vanishes with time, and become a disadvantage after about four months. This is more evident by estimating the number of cases \textit{per capita} from Eq.~(\ref{eq:scaling_cases}), that is, $Y_c/P\sim P^{\beta_c-1}$. Similarly, we can estimate the number of deaths \textit{per capita} from Eq.~(\ref{eq:scaling_deaths}), yielding $Y_d/P\sim P^{\beta_d-1}$. Thus, we expect the number of COVID-19 cases or deaths \textit{per capita} to decrease with the city population if $\beta_c<1$ and $\beta_d<1$; conversely, these \textit{per capita} numbers are expected to increase with the city population if $\beta_c>1$ and $\beta_d>1$. For instance, because $\beta_c\approx0.77$ and $\beta_d\approx0.85$ after 75 days since the first two daily cases or deaths, the number of cases and deaths \textit{per capita} decreases with population as $Y_c/P\sim P^{-0.23}$ and $Y_c/P\sim P^{-0.15}$. At those particular values of $t_c$ and $t_d$, an $1\%$ rise in the population is associated with a $\approx0.23\%$ decrease in the incidence of COVID-19 cases and $\approx0.15\%$ reduction in the incidence of deaths. In a concrete example for $t_c=t_d=75$~days, we expect a metropolis such as S\~ao Paulo (with $\approx12$ million people) to have $\approx 54\%$ less cases and $\approx 39\%$ less deaths \textit{per capita} than a medium-sized city such as Maring\'a/PR (with $\approx420$ thousand people, $\approx1/30$ of S\~ao Paulo), which in turn is expected to have $\approx 41\%$ less cases and $\approx 29\%$ less deaths \textit{per capita} than a small-sized city such as Parana\'iba/MS (with $\approx42$ thousand people, $\approx1/10$ of Maring\'a). 

However, both scaling exponents increase with time, such that this urban advantage vanishes and become a disadvantage during the long course of the pandemic. By considering our latest estimates for the scaling exponents, we find $\beta_c\approx1.04$ ($t_c=144$~days) and $\beta_d\approx1.12$ ($t_d=120$~days). Thus, at these particular values of $t_c$ and $t_d$, we expect the number of cases \textit{per capita} to slightly increase with population ($Y_c/P\sim P^{0.04}$) and the number of fatalities \textit{per capita} to increase with population as $Y_d/P\sim P^{0.12}$. Thus, for $\beta_d\approx1.12$ at $t_d=120$~days, we expect a metropolis such as S\~ao Paulo ($\approx12$ million people) to have $\approx 50\%$ more deaths \textit{per capita} than Maring\'a/PR ($\approx420$ thousand people), which in turn is expected to have $\approx 32\%$ more deaths \textit{per capita} than Parana\'iba/MS ($\approx42$ thousand people). Figures~8-14 in S1~Appendix show that the scaling relations for number of cases and deaths \textit{per capita} support the previous discussions.

The latest estimates of $\beta_c$ found for cases of COVID-19 are smaller than those reported for the 2009 H1N1 Pandemic in Brazil ($\beta_c\approx1.2$) and HIV in Brazil and United States ($\beta_c\approx1.4$)~\cite{rocha2015non}. Similarly to what we observe for the cases of COVID-19, the allometric exponent for HIV cases in Brazil was initially sub-linear during the 1980s, became super-linear after the 1990s, and started to approach a super-linear plateau after the 2000s~\cite{rocha2015non}. However, the evolution of the allometry for HIV has been much slower than what we have observed for the COVID-19. Another interesting point reported by Rocha, Thorson, and Lambiotte~\cite{rocha2015non} is that the number of H1N1 cases in Brazil started to scale linearly with city population in 2010 (one year after the first outbreak). These authors also argue that this reduction in the scaling exponent possibly reflects a better response for the spread of H1N1 after the pandemic outbreak. If the behavior observed in the 2009 H1N1 Pandemic generalizes (at least in part) for the current COVID-19 pandemic, we would expect a decrease in values of $\beta_c$ in the future. The lastest estimates of $\beta_d$ for COVID-19 deaths are larger than those reported for diabetes ($\beta_d\approx0.8$), heart attack ($\beta_d\approx1$) and cerebrovascular accident ($\beta_d\approx1$) in Brazil after the 2000s~\cite{rocha2015non}. Conversely, scaling exponents related to disease mortality in Brazil displayed a decreasing trend with time, and values as high as $1.25$ were observed for diabetes in 1996 ($\beta_d\approx1.22$) and heart attack in 1981 ($\beta_d\approx1.25$)~\cite{rocha2015non}. The convergence of these exponents to linear or sub-linear regimes may reflect the increasing access to medical facilities in urban areas~\cite{rocha2015non}.

Based on currently available data (Figure~\ref{fig:2}), it is hard to confidently assert whether the values of $\beta_c$ and $\beta_d$ will remain larger than one during the long-term course of the pandemic. However, the persistence of this behavior indicates large cities are likely to be more affected at the end of the COVID-19 outbreak. Part of this behavior may be due to large cities testing for COVID-19 proportionally more than small ones. Results for the United States indicate that more rural states have lower testing rates and detect disproportionately fewer cases of COVID-19~\cite{souch2020commentary}. As Brazilian cities are likely to suffer from this bias, we would expect a decrease in the scaling exponent $\beta_c$ after the observed increasing trend depending on the magnitude of this effect (that is, as small cities increase their testing capabilities, their number of cases tend to increase and bend the scaling law downwards).

On the other hand, it is clearer that large cities were proportionately less affected during the initial months (since the first two daily cases or deaths) of the pandemic. We believe there are at least two possible explanations for this behavior. First, it may reflect an ``increasing urban advantage'' where the larger the city, the more access to medical facilities and so the chance of receiving more appropriate treatment against the coronavirus disease. A second cause can be associated with age demographic changes with the city population; specifically, a smaller proportion of older adults at high risk for severe illness and death from COVID-19 leads to a reduced number of deaths \textit{per capita}. Another possibility is that the strategies and policy responses of large and small cities to COVID-19 are different, which in turn may lead to different efficiency in containing the pandemic. These responses are highly heterogeneous at the national level~\cite{hsiang2020effect,gao2020quantifying} as well as among counties in the United States~\cite{brandtner2020creatures}. Among these three possibilities, we did not explore the possible effects of different city strategies against the COVID-19, but in light of the findings for the United States~\cite{brandtner2020creatures}, this effect is likely to play an important role in the Brazilian case and may deserve further investigation.

\begin{figure*}
    \centering
    \includegraphics[width=\textwidth]{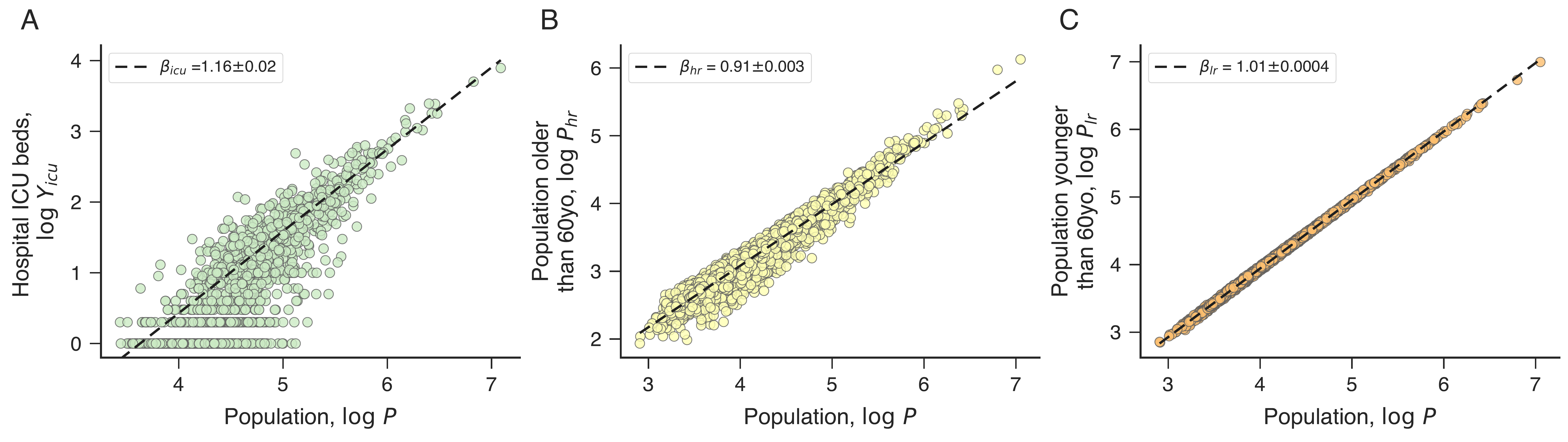}
    \caption{\textbf{Urban scaling of ICU beds, high-risk and low-risk populations.} (A) Relationship between the number of ICU beds ($Y_{icu}$) and the city population ($P$) on logarithmic scale. We observe that the number of ICU beds scales super-linearly with city size ($\beta_{icu}=1.16\pm0.02$), indicating an urban advantage for health coverage. (B) Relationship between the high-risk population ($P_{hr}$, defined as adults older than 60 years) and the city population ($P$) on logarithmic scale. The high-risk population scales sub-linearly ($\beta_{hr}=0.910\pm0.003$), showing that large cities tend to have smaller fractions of elderly than small cities. (C) Relationship between low-risk population ($P_{lr}$, defined as adults younger than 60 years) and the city population ($P$) on logarithmic scale. We note that the low-risk population scales almost linearly ($\beta_{hr}=1.0100\pm0.0004$) with city size. The behavior of these three quantities partially explains the initial decrease of number of deaths \textit{per capita} with population ($\beta_d<1$ for $t_d\lesssim100$ days). See Figure~15 in S1~Appendix for the scaling relations involving \textit{per capita} quantities.}
    \label{fig:3}
\end{figure*}

To test for an increasing urban advantage for the treatment of COVID-19 during the initial spread of the disease, we investigate the scaling relation between the number of hospital intensive care unit (ICU) beds and city population. Because critically ill patients frequently require mechanical ventilation~\cite{grasselli2020critical,arabi2020covid}, the number of ICU beds has proved to be crucial for the treatment of COVID-19. Figure~\ref{fig:3}A shows the allometric relationship between the number of ICU beds from private and public health systems ($Y_{icu}$, as of April 2020) and the population, where a super-linear relationship emerges with scaling exponent $\beta_{icu}\approx1.16$. The super-linear scaling of ICU beds indicates that large Brazilian cities are better structured to deal with critically ill patients, which in turn may partially explain the reduction of deaths \textit{per capita} with the city size during the initial three-four months since the first two daily deaths. It is worth noting that the Brazilian Public Unified Health System (Sistema \'Unico de Sa\'ude -- SUS) is decentralized and composed of ``health regions'', contiguous groups of cities usually formed by a large city and its neighboring cities~\cite{castro2019brazil}. Cities within the same health region may share medical services, which may in turn partially explain the reduction of the structural advantages of large urban areas during the long-term course of the pandemic.

We have also investigated how age demographic distribution changes with city population. Estimates have shown that the case fatality rate of COVID-19 is substantially higher in people aged more than 60 years (0.32\% for those younger than 60 years versus 6.5\% for those older than 60 years~\cite{verity2020estimates}). Thus, the age demographic of cities represents an important factor for the number of deaths caused by COVID-19. Figures~\ref{fig:3}B and \ref{fig:3}C show how the number of people older ($P_{hr}$, the high-risk population) and younger ($P_{lr}$, the low-risk population) than 60 years change with the total population ($P$). We note that the high-risk population increases sub-linearly with city size with an exponent $\beta_{hr}\approx0.91$, while the low-risk population scales linearly ($\beta_{lr}\approx1$) with city size. This result shows that large cities have a lower prevalence of adults older than 60 years, such that a $1\%$ increase in city population is associated with a $0.91\%$ rise in the high-risk population. In a more concrete example, we expect a city with one million people to have proportionally $\approx19\%$ fewer adults older than 60 years when compared with a city of 100 thousand inhabitants. Thus, a low prevalence of elderly in large urban areas may also partially explain the initial reduction of the number of deaths \textit{per capita} with the increase of city population.

In addition to addressing the urban scaling of cases and deaths of COVID-19, we have investigated associations between the growth rates of cases and deaths and the city population (Figures~16-22 in S1~Appendix). As mentioned, the work of Stier, Berman, and Bettencourt~\cite{stier2020covid} shows that the initial growth rates of COVID-19 cases in metropolitan areas of the United States scale as a power-law function of the population with an exponent between $0.11$ and $0.20$. By using our data and as detailed in Methods, we have estimated the growth rates of cases ($r_c$) and deaths ($r_d$) for Brazilian cities. In agreement with the United States case, our results also indicate that COVID-19 cases initially grow faster in large cities (Figure~23 in S1~Appendix), such that $r_c\sim P^{\beta_{r_c}}$ with $\beta_{r_c}$ between $\approx0.1$ and $\approx0.3$ during the first three months ($t_c\lesssim90$, Figure~23 in S1~Appendix). We also found similar behavior for the growth rate in the number of deaths $r_d$, where a power-law relation $r_d\sim P^{\beta_{r_d}}$ is a reasonable description for the empirical data with a scaling exponent $\beta_{r_d}$ between $\approx0.1$ and $\approx0.5$ during the first three months ($t_d\lesssim90$, Figure~23 in S1~Appendix).

\begin{figure*}
    \centering
    \includegraphics[width=0.8\textwidth]{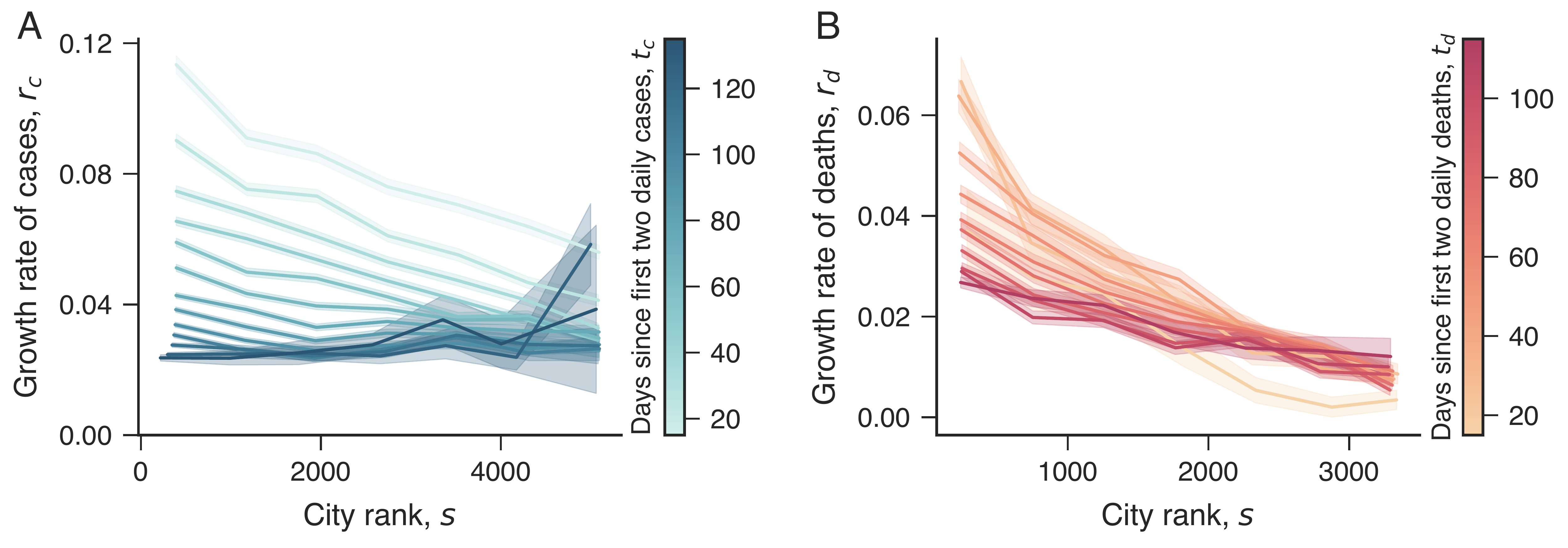}
    \caption{\textbf{Association between growth rates and city size.} (A) Relationship between the growth rate of COVID-19 cases ($r_c$) and the city rank ($s$). The different curves show the average values $r_c$ versus $s$ for different number of days since the first two daily cases ($t_c$, as indicated by the color code). (B) Relationship between the growth rate of deaths by COVID-19 ($r_d$) and the city rank ($s$). The different curves show the average values $r_d$ versus $s$ for different number of days since the first two daily deaths ($t_d$, as indicated by the color code).}
    \label{fig:4}
\end{figure*}

The growth rate depicts a more instantaneous picture of the COVID-19 spreading process, and its association with size may change during the long-term evolution of the pandemic. These changes may reflect the different actions taken by each city to face the COVID-19 pandemic and other particularities affecting the COVID-19 spreading. For the spreading of COVID-19 in the United States, Heroy~\cite{heroy2020metropolitan} has reported that large cities appear to enter in an exponential spreading regime earlier than small ones. To better investigate these possibilities in our data, we have estimated the average relationship between the growth rate of cases ($r_c$) and deaths ($r_d$) and the city rank $s$ ($s=1$ represents the largest city in data, $s=2$ the second-largest, and so on) at different periods. Figure~\ref{fig:4}A shows the results for the growth rates in the number of cases ($r_c$). In agreement with the power-law association between $r_c$ and the city population (Figures~16-22 in S1~Appendix), we note that lower values of the city rank $s$ are associated with higher growth rates $r_c$ in the initial days since the first two daily cases. However, as time goes by, the growth rate of cases starts to decrease in large cities (low-rank values) and to increase in small ones (high-rank values). This result appears to agree with the findings of Heroy~\cite{heroy2020metropolitan} in the sense that there is a delay in the emergence of high growth rates of cases between large and small cities. Figure~\ref{fig:4}B shows the same analysis for growth rate in the number of deaths $r_d$. While we also observe a decrease in $r_d$ for large cities and increase for small ones, the differences in $r_d$ are less pronounced than in $r_c$. These findings also emerge when investigating the scaling exponents associated with the growth rates of cases ($\beta_{r_c}$) and deaths ($\beta_{r_d}$). The results of Figure~23 in S1~Appendix show that these exponents start to decrease around $t_c\approx t_d\approx100$ days and become negative in our latest estimates. It is worth remembering that the time $t_c$ (or $t_d$) is measured in days since the first two daily cases (or first two daily deaths) for each city; thus, the results of Figure~\ref{fig:4} do not reflect delays in the emergence of the first case in each city. 

\section*{Discussion}
We have studied scaling relations for the number of COVID-19 cases and deaths in Brazilian cities. Similarly to what happens for other diseases, we found the number of cases and deaths to be power-law related to the city population. During the initial three-four months since the first two daily cases or deaths, we found a sub-linear association between cases and deaths by COVID-19, meaning that the \textit{per capita} numbers of cases and deaths tend to decrease with population in this initial stage of the pandemic. We believe this behavior can be partially explained by an ``increasing urban advantage'' where large cities have proportionally more ICU beds than small ones. In addition, changes in age demography with city size show that large cities have proportionally less elderly people who are at high risk of developing severe illness and dying from COVID-19. This may also partially explain the initial reduction of fatalities \textit{per capita} with the city population. In addition, we have argued that the strategies and policy responses of large and small cities to COVID-19 may also be different and lead to different efficiency in containing the pandemic.

However, we found that this ``urban advantage'' vanishes in the long-term course of the pandemic, such that the association between cases and deaths by COVID-19 with population becomes super-linear in our latest estimates since the first two daily cases or deaths. Thus, the persistence of this pattern indicates that large cities are expected to be proportionally more affected at the end of the COVID-19 pandemic. This result is in line with the findings for other infectious diseases~\cite{rocha2015non,antonio2014growth} and probably reflects the existence of a higher degree of interaction between people in large cities~\cite{bettencourt2013origins,schlapfer2014scaling}. Because social distancing is currently the only available measure to mitigate the impact of COVID-19, our results suggest that large cities may require more severe degrees of social distancing policies. 

In agreement with the results for metropolitan areas in the United States~\cite{stier2020covid}, we have found that large cities usually display higher growth rates in the number of cases during the initial spread of the COVID-19. However, our results also show that these growth rates tend to decrease in large cities and to increase in small ones in the long-term course of the pandemic. This behavior suggests the existence of a delay in the emergence of high growth rates between large and small cities. Similar behavior was also found in the United States~\cite{heroy2020metropolitan}, where large cities appear to enter an exponential growth regime earlier than small towns. The existence of this delay suggests that the initial slow-spreading pace of the COVID-19 in small cities is likely to be a transient behavior.

Together with the recent findings of Stier-Berman-Bettencourt~\cite{stier2020covid} and Heroy~\cite{heroy2020metropolitan} for the United States, as well as those of Cardoso and Gon\c{c}alves~\cite{cardoso2020urban} for United States, Brazil and Germany, our results suggest that social distancing policies and other actions against the pandemic should take into account the non-linear effects of city size on the spreading of the COVID-19.

\section*{Methods}

\subsection*{Data}
The primary data set used in this work was collected from the \texttt{brasil.io} API~\cite{brasil_io}. This API retrieves information from COVID-19 daily reports published by the Health Offices of each of the 27 Brazilian federations (26 states and one federation district) and makes it freely available. This data set comprises information about the cumulative number of cases and deaths of COVID-19 from 25 February 2020 (date of the first case in Brazil) until 12 August 2020 (date of our last update) for all Brazilian cities reporting at least one case of COVID-19. The \texttt{brasil.io} API also provides population data of Brazilian cities, which in turn relies on population estimates for the year 2019 released by the Brazilian Institute of Geography and Statistics (IBGE). There is a total of 5,507 Brazilian cities with at least one reported case of COVID-19 on 12 August 2020, corresponding to 98.9\% of the country's total number of cities. In addition, 3,892 cities suffered casualties from this disease, representing 69.9\% of the total. To ensure that our estimates rely on at least 50 cities, we consider a suitable upper threshold for the time series length (Figure~24 in S1~Appendix). The data about age demographics refer to the latest Brazilian census that took place in 2010, while the data about the number of ICU beds are from April 2020. These two data sets are maintained and made freely available by the Department of Informatics of the Brazilian Public Health System (DATASUS)~\cite{datasus}.

\subsection*{Fitting urban scaling laws}
Urban scaling~\cite{bettencourtgrowth2007} usually refers to a power-law association between a city property $Y$ and the city population $P$, and it is expressed by
\begin{equation}\label{eq:scaling}
    Y = Y_0 P^\beta\,,
\end{equation}
where $Y_0$ is a constant and $\beta$ is the urban scaling exponent. Equation~(\ref{eq:scaling}) can be linearized by taking the logarithmic on both sides, that is,
\begin{equation}\label{eq:scaling_lin}
    \log Y = \log Y_0 + \beta \log P\,,
\end{equation}
where $\log Y$ and $\log P$ are the dependent and independent variables of the corresponding linear relationship between $\log Y$ and $\log P$. We have estimated the power-law exponents in Eq.~(\ref{eq:scaling}) by using the probabilistic approach of Leit\~ao \textit{et al.}~\cite{leitao2016scaling}. Specifically, we have found the probabilistic model with lognormal fluctuations and where the fluctuations in $\log Y$ are independent of $P$ to be the best description of our data in the majority of scaling laws. Thus, we assume these lognomal fluctuations in all adjusting procedures in order to estimate the values of $\beta$. It is worth mentioning that this maximum-likelihood estimate for scaling exponents is analogous to the one obtained via usual least-squares with the log-transformed variables ($\log Y$ versus $\log P$).

\subsection*{Logarithmic growth rates of cases and deaths}
Let us consider that $x_t$ ($t=1,\dots, n$) represents the cumulative number of cases ($Y_c$) or the cumulative number of deaths ($Y_d$) for COVID-19 in a given city at time $t$ (number of days since first case $t_c$ or death $t_d$). The logarithmic growth rate $r_t$ at time $t$ is defined as
\begin{equation}\label{eq:lograte}
    r_t = \log ({x_{t}}/{x_{t-\tau}})/\tau~~(t=\tau,\tau+1,\dots,n)\,
\end{equation}
where $\tau$ is a time delay. If we assume the numbers of cases or deaths to initially increase exponentially ($x_t \sim e^{r t}$, where $r$ is the exponential growth rate), $r_t$ represents an estimate for the growth rate of this initial exponential behavior ($r$). We have estimated $r_t$ for the number of cases ($r_c$) and deaths ($r_d$) up to values of $t_c$ and $t_d$ ensuring a sample size of at least 50 cities for the allometric relations between these growth rates and the city population (Figure~24 in S1~Appendix). All results in the main text were obtained for $\tau=14$ but our discussion is robust for $\tau$ between $9$ and $21$ days (Figures~25-38 in S1~Appendix).

\section*{Acknowledgements}
This research was supported by Coordena\c{c}\~ao de Aperfei\c{c}oamento de Pessoal de N\'ivel Superior (CAPES) and Conselho Nacional de Desenvolvimento Cient\'ifico e Tecnol\'ogico (CNPq). H.V.R. thanks for the financial support of CNPq (Grant Nos. 407690/2018-2 and 303121/2018-1). M.P. acknowledges financial support of the Slovenian Research Agency (Grant Nos. J4-9302, J1-9112, and P1-0403). 

\section*{Data availability}
All data supporting the findings of this study are freely available as detailed in the main text.

\section*{Author contributions statement}
H.V.R, A.S.S., J.S., M.P., and Q.S.H. designed research, performed research, analyzed data, and wrote the paper.

\section*{Additional information}

\noindent \textbf{Competing interests:} The authors declare that they have no conflict of interest. 

\section*{S1 Appendix}
Supplementary Figures (1-38) supporting the robustness of our findings against different reference points for synchronizing the time series of cases and deaths among cities, different time delays used for estimating the growth rates, and other additional figures.

\bibliographystyle{naturemag}
\bibliography{city_size_covid}
\pagebreak

\setcounter{page}{1}
\setcounter{figure}{0}
\makeatletter
\renewcommand{\figurename}{Figure}
\renewcommand{\tablename}{Table}
\renewcommand{\thefigure}{\@arabic\c@figure}
\renewcommand{\thetable}{\@arabic\c@table}
\clearpage

\onecolumngrid
\thispagestyle{empty}
\begin{center}
\large{S1 Appendix for}\\
\vskip1pc
\large{\bf City size and the spreading of COVID-19 in Brazil}\\
\vskip1pc
\normalsize{Haroldo\ V.\ Ribeiro, Andre\ S.\ Sunahara, Jack Sutton, Matja{\v z} Perc, and Quentin S. Hanley}\\
\vskip1pc
\normalsize{PLOS ONE, 2020}
\end{center}


\begin{figure*}[!ht]
    \centering
    \includegraphics[width=1\textwidth]{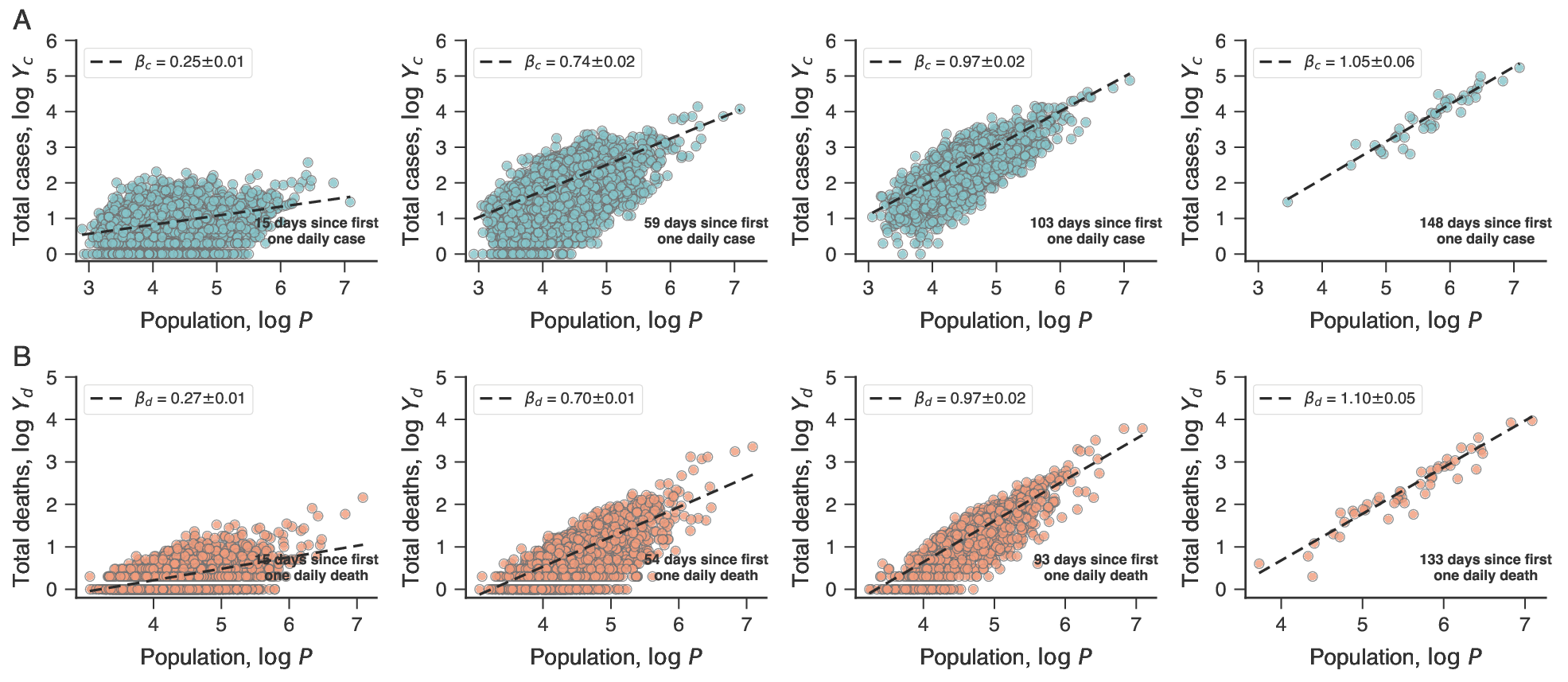}
    \caption{\textbf{Urban scaling relations of COVID-19 cases and deaths under different choices of values for the number of daily cases or daily deaths as reference points.} The same plots of Figure~1 in the main text but considering the first one daily case and first one daily death as reference points.}
    \label{sfig:fig1_1}
\end{figure*}
\clearpage

\begin{figure*}
    \centering
    \includegraphics[width=1\textwidth]{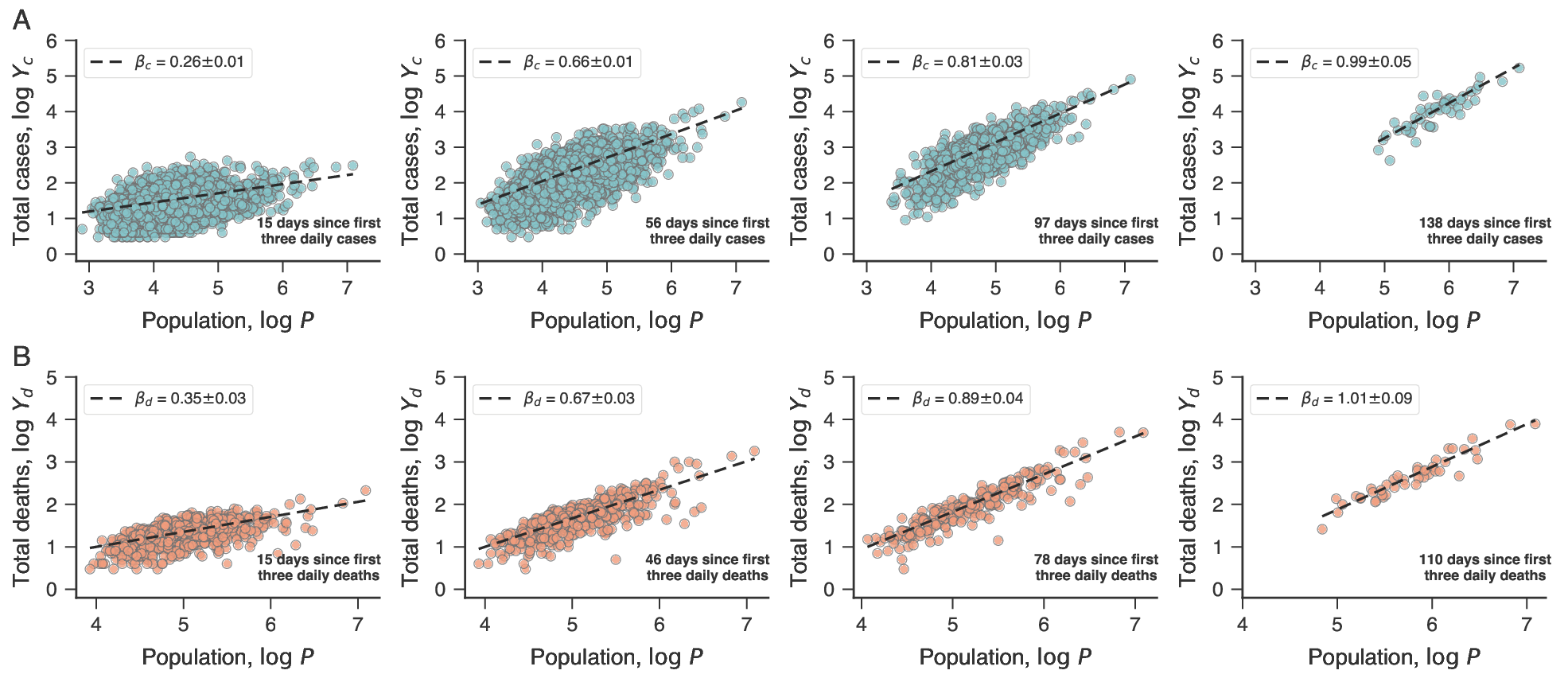}
    \caption{\textbf{Urban scaling relations of COVID-19 cases and deaths under different choices of values for the number of daily cases or daily deaths as reference points.} The same plots of Figure~1 in the main text but considering the first three daily cases and the first three daily deaths as reference points.}
    \label{sfig:fig1_3}
\end{figure*}
\clearpage

\begin{figure*}
    \centering
    \includegraphics[width=1\textwidth]{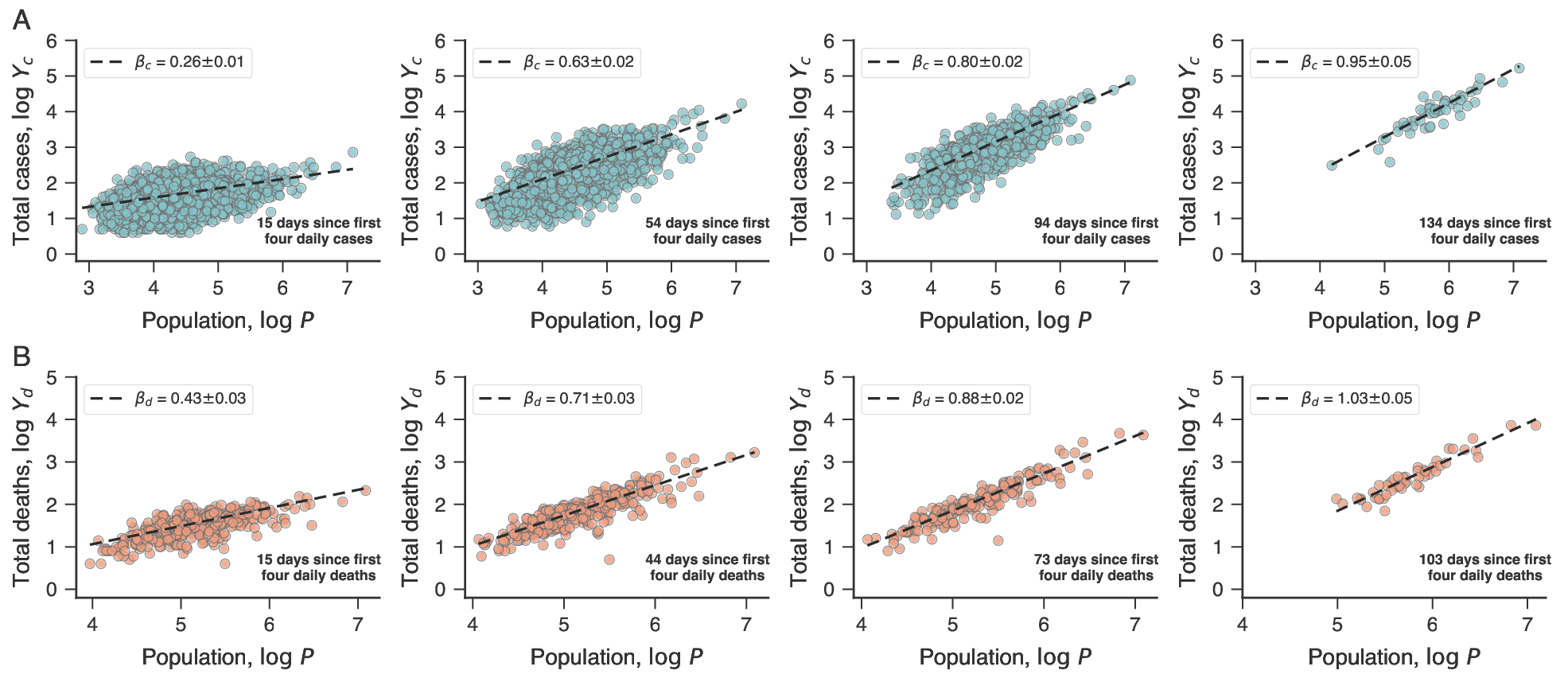}
    \caption{\textbf{Urban scaling relations of COVID-19 cases and deaths under different choices of values for the number of daily cases or daily deaths as reference points.} The same plots of Figure~1 in the main text but considering the first four daily cases and the first four daily deaths as reference points.}
    \label{sfig:fig1_4}
\end{figure*}
\clearpage

\begin{figure*}
    \centering
    \includegraphics[width=1\textwidth]{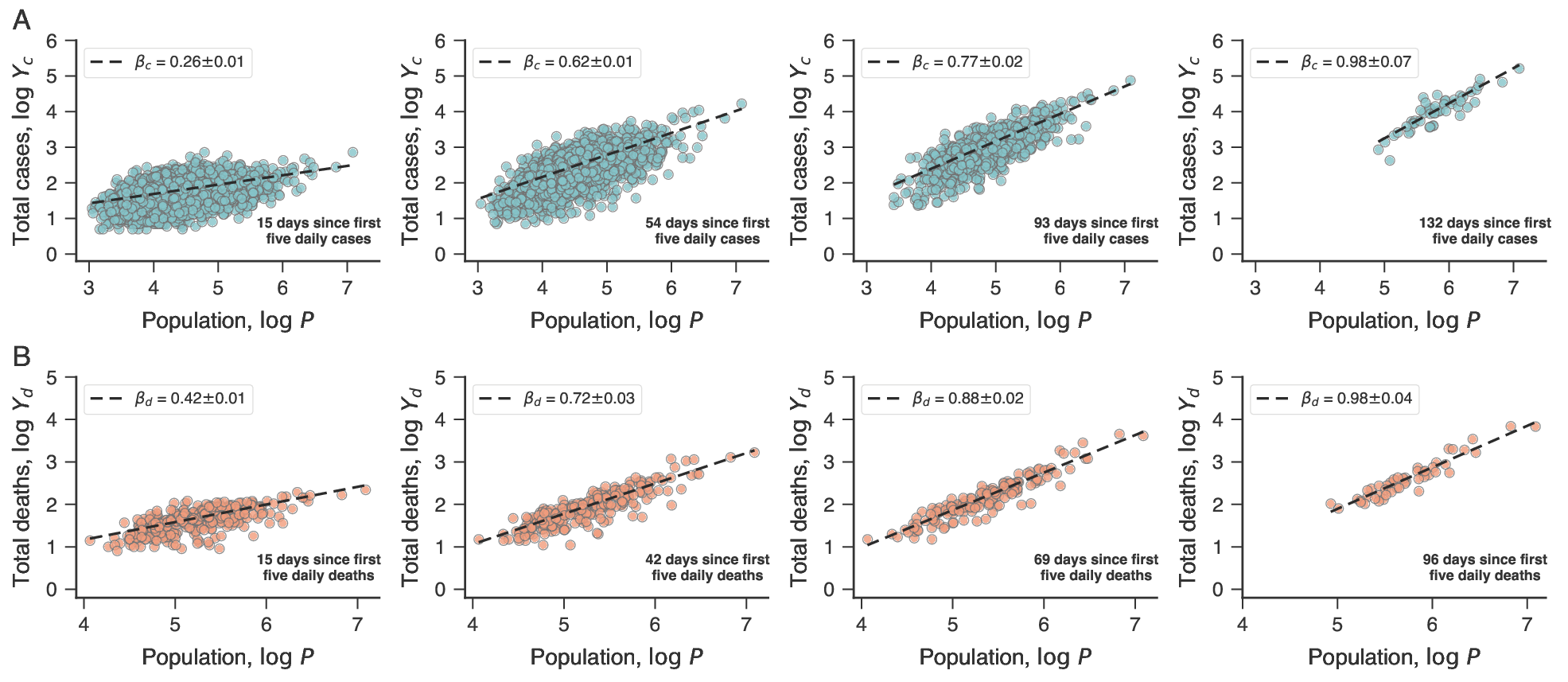}
    \caption{\textbf{Urban scaling relations of COVID-19 cases and deaths under different choices of values for the number of daily cases or daily deaths as reference points.} The same plots of Figure~1 in the main text but considering the first five daily cases and the first five daily deaths as reference points.}
    \label{sfig:fig1_5}
\end{figure*}
\clearpage

\begin{figure*}
    \centering
    \includegraphics[width=1\textwidth]{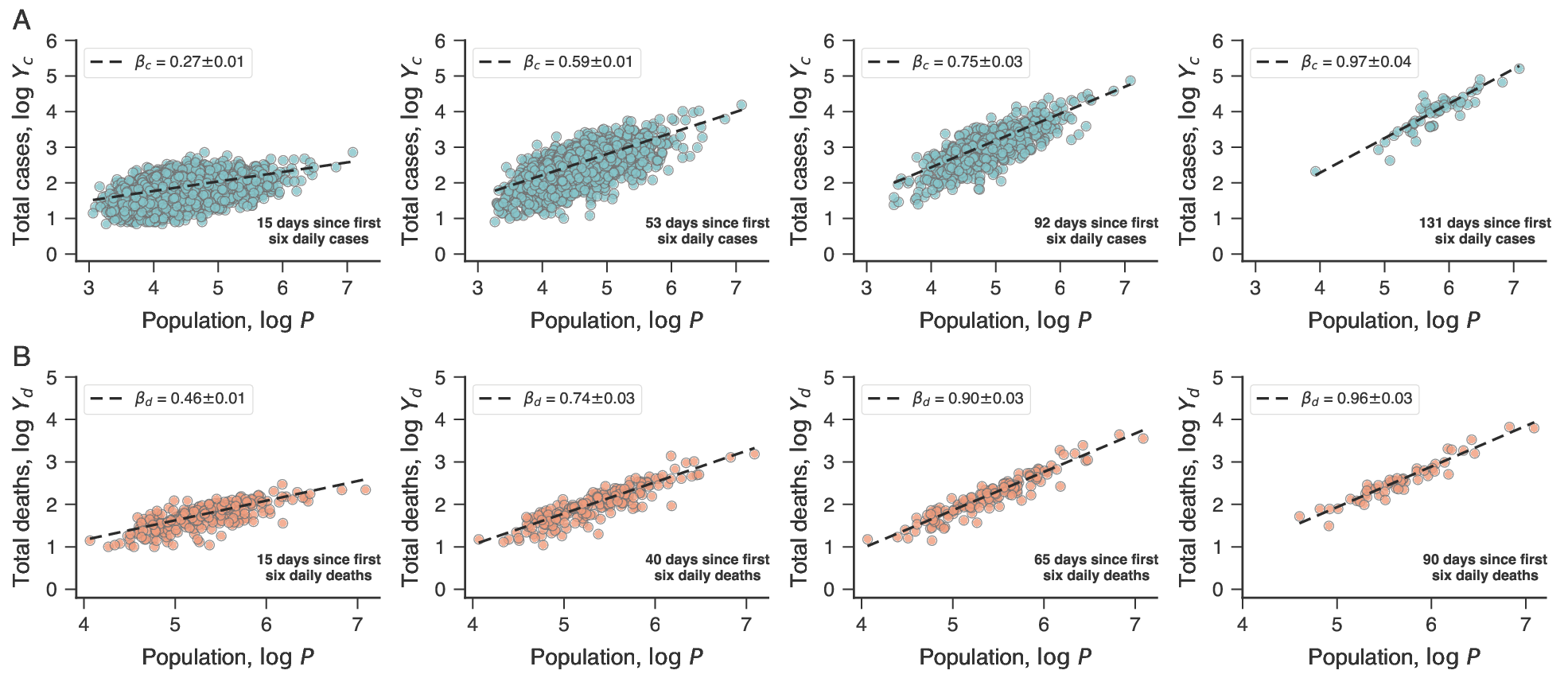}
    \caption{\textbf{Urban scaling relations of COVID-19 cases and deaths under different choices of values for the number of daily cases or daily deaths as reference points.} The same plots of Figure~1 in the main text but considering the first six daily cases and the first six daily deaths as reference points.}
    \label{sfig:fig1_6}
\end{figure*}
\clearpage

\begin{figure*}
    \centering
    \includegraphics[width=1\textwidth]{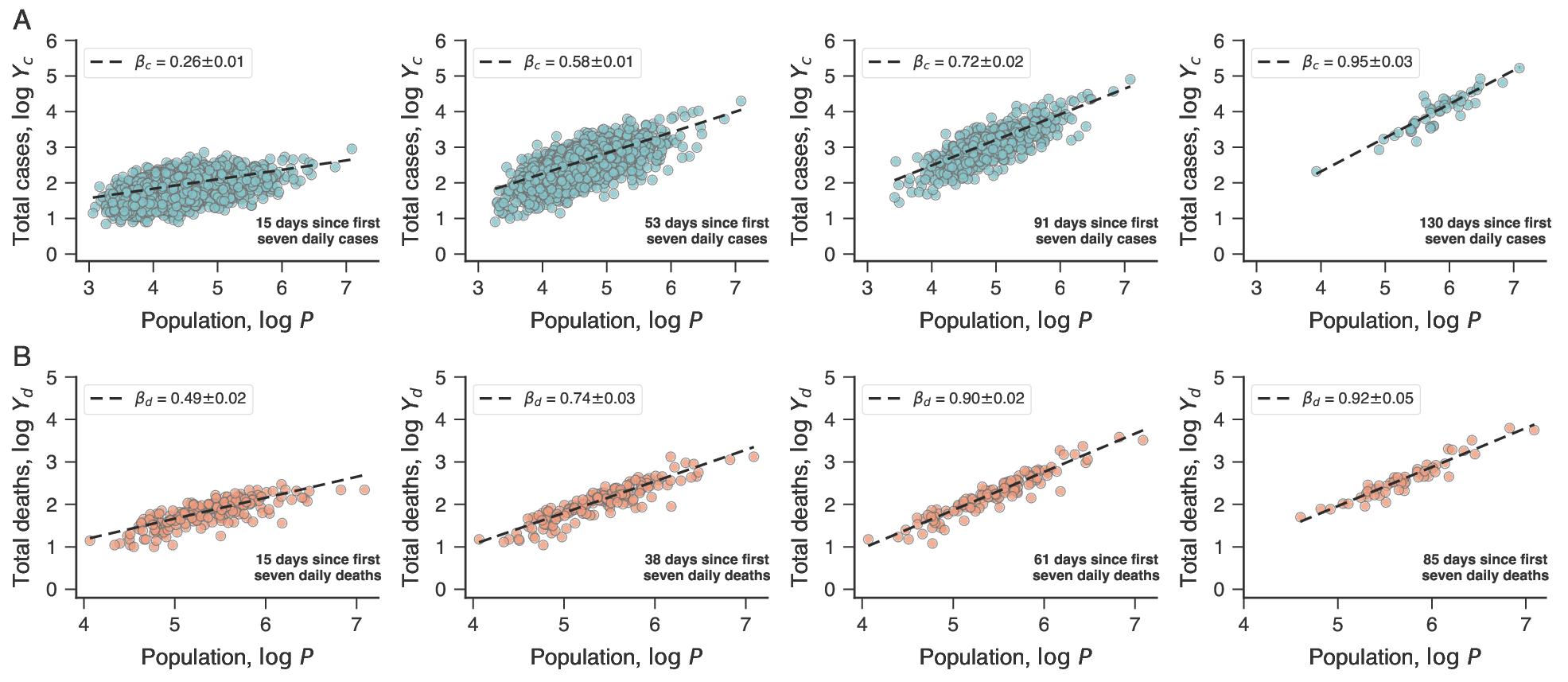}
    \caption{\textbf{Urban scaling relations of COVID-19 cases and deaths under different choices of values for the number of daily cases or daily deaths as reference points.} The same plots of Figure~1 in the main text but considering the first seven daily cases and the first seven daily deaths as reference points.}
    \label{sfig:fig1_7}
\end{figure*}
\clearpage

\begin{figure*}
    \centering
    \includegraphics[width=0.44\textwidth]{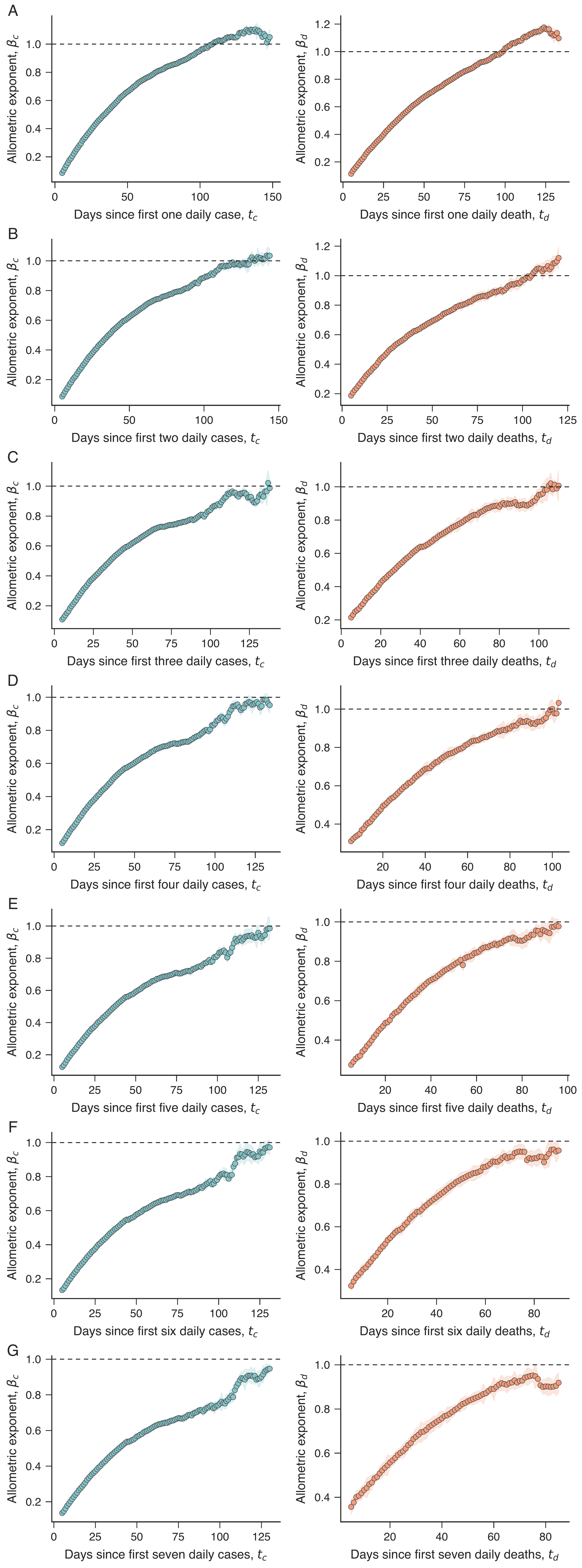}
    \caption{\textbf{Time dependence of the scaling exponents for COVID-19 cases and deaths under different choices of values for the number of daily cases or daily deaths as reference points.} Panels (A)-(G) show the dependence of the exponents $\beta_c$ (left) and $\beta_d$ (right) on $t_c$ and $t_d$ when considering the first 1-7 daily cases and the first 1-7 daily deaths as reference points. The shaded regions stand for standard errors, and the horizontal dashed lines represent $\beta_{c}=\beta_{d}=1$. We note that the behavior observed for large numbers of the reference points appear to follow the behavior of small ones in the long-term course of the pandemic.}
    \label{sfig:fig2}
\end{figure*}
\clearpage

\begin{figure*}
    \centering
    \includegraphics[width=1\textwidth]{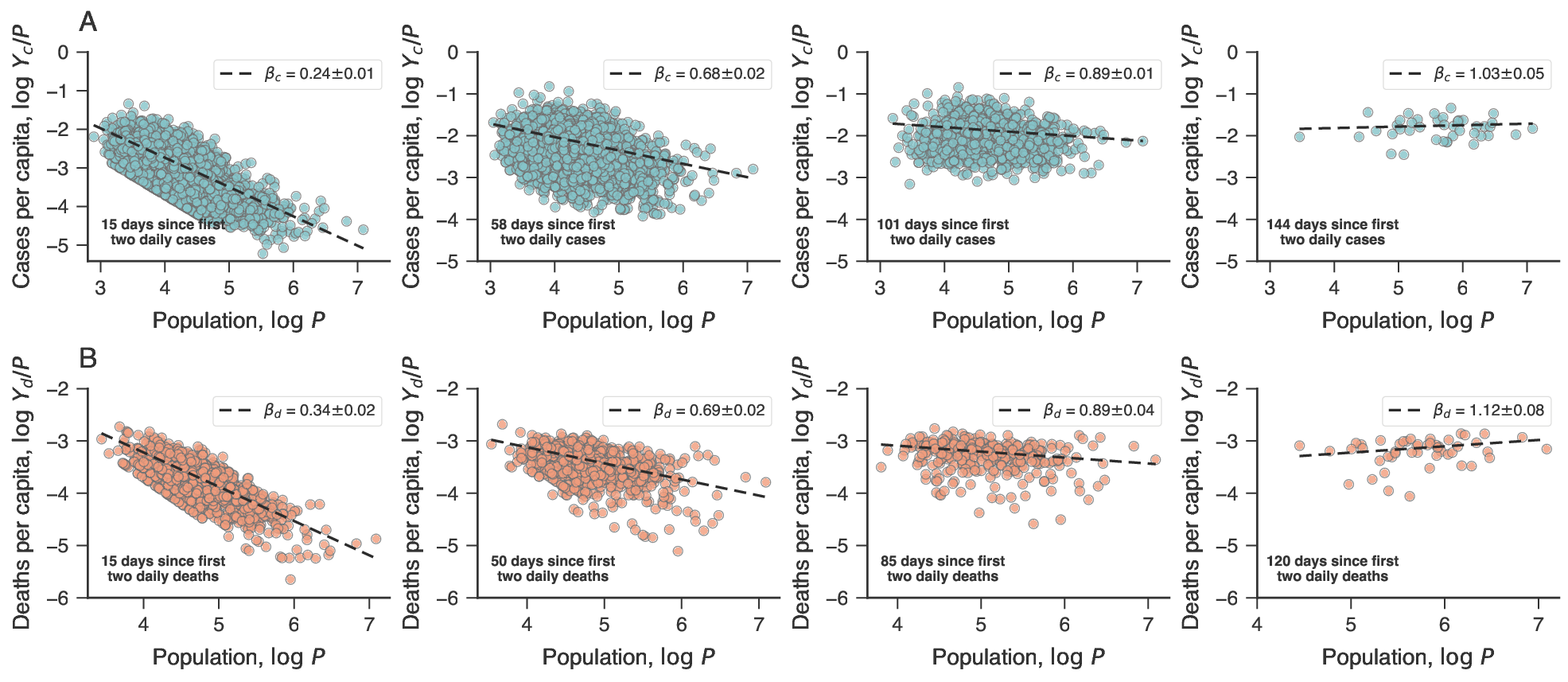}
    \caption{\textbf{Urban scaling relations of COVID-19 cases and deaths \textit{per capita}.} (A) Relationship between the total of confirmed cases \textit{per capita} of COVID-19 ($Y_c/P$) and city population ($P$) on logarithmic scale. (B) Relationship between the total of deaths \textit{per capita} caused by COVID-19 ($Y_d$) and population ($P$) of Brazilian cities (on logarithmic scale). Panels show scaling relations on a particular day after the first two daily cases or the first two daily deaths. The dashed lines are the scaling relations with exponents indicated in each plot ($\beta_c-1$ for cases \textit{per capita} and $\beta_d-1$ for deaths \textit{per capita}).}
    \label{sfig:fig1_2_per}
\end{figure*}
\clearpage

\begin{figure*}
    \centering
    \includegraphics[width=1\textwidth]{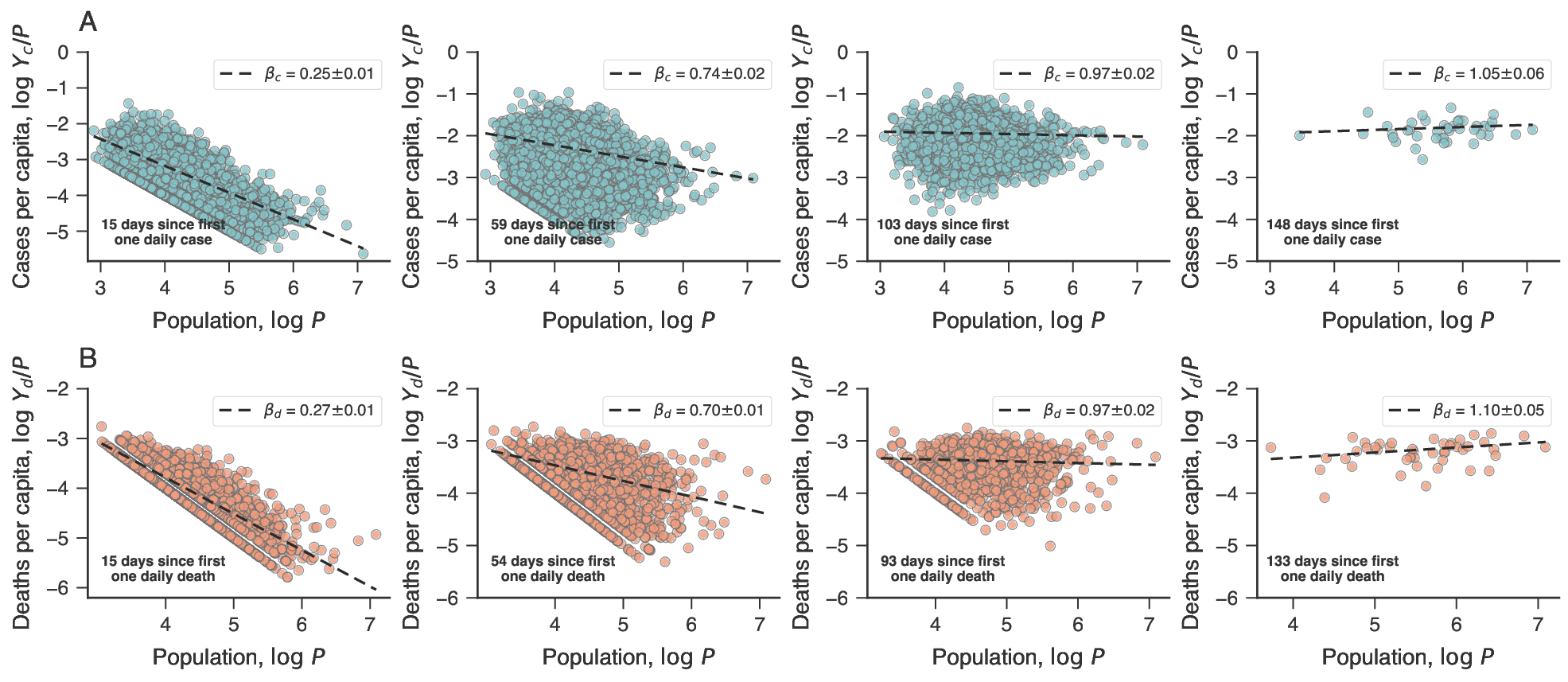}
    \caption{\textbf{Urban scaling relations of COVID-19 cases and deaths \textit{per capita}.} The same as Figure~8 in this Appendix, but considering the first one daily case and the first one daily death as reference points.}
    \label{sfig:fig1_1_per}
\end{figure*}
\clearpage

\begin{figure*}
    \centering
    \includegraphics[width=1\textwidth]{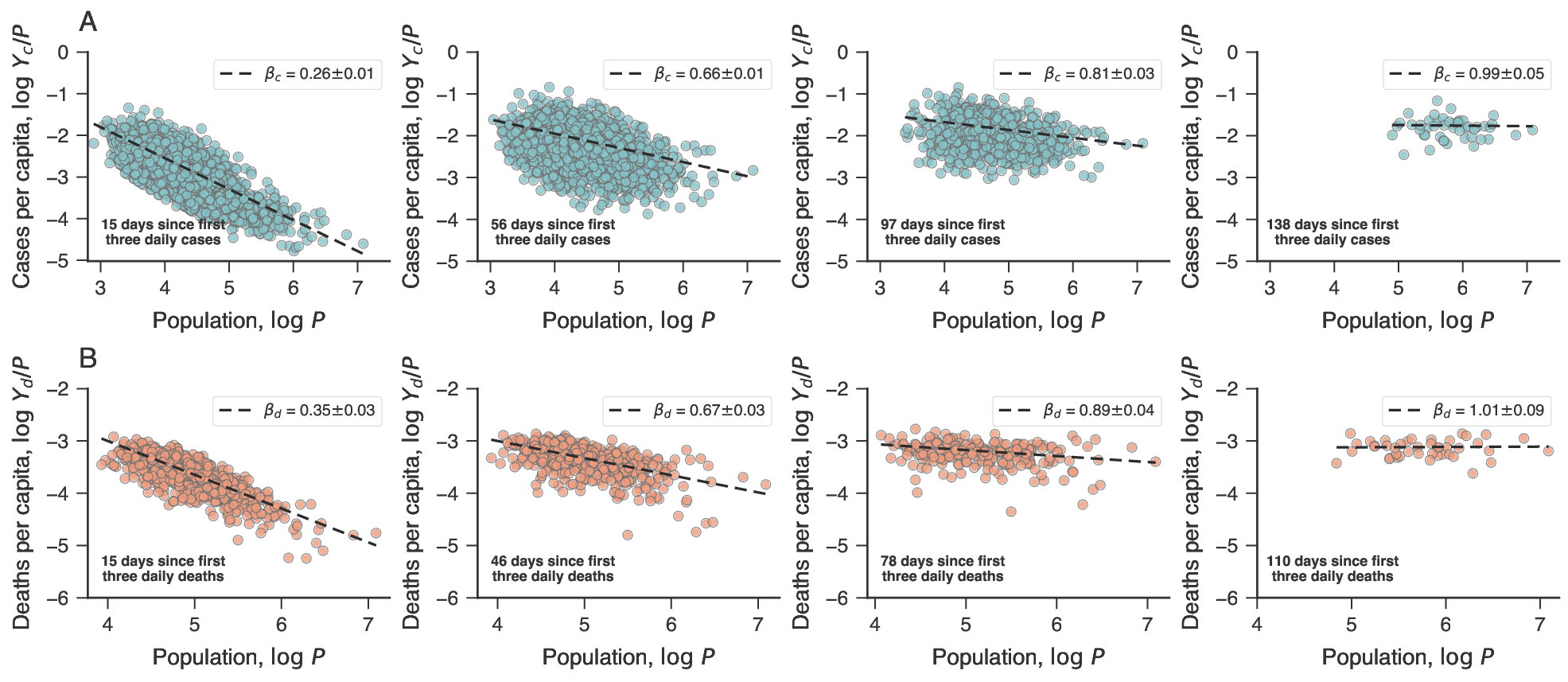}
    \caption{\textbf{Urban scaling relations of COVID-19 cases and deaths \textit{per capita}.} The same as Figure~8 in this Appendix, but considering the first three daily cases and the first three daily deaths as reference points.}
    \label{sfig:fig1_3_per}
\end{figure*}
\clearpage

\begin{figure*}
    \centering
    \includegraphics[width=1\textwidth]{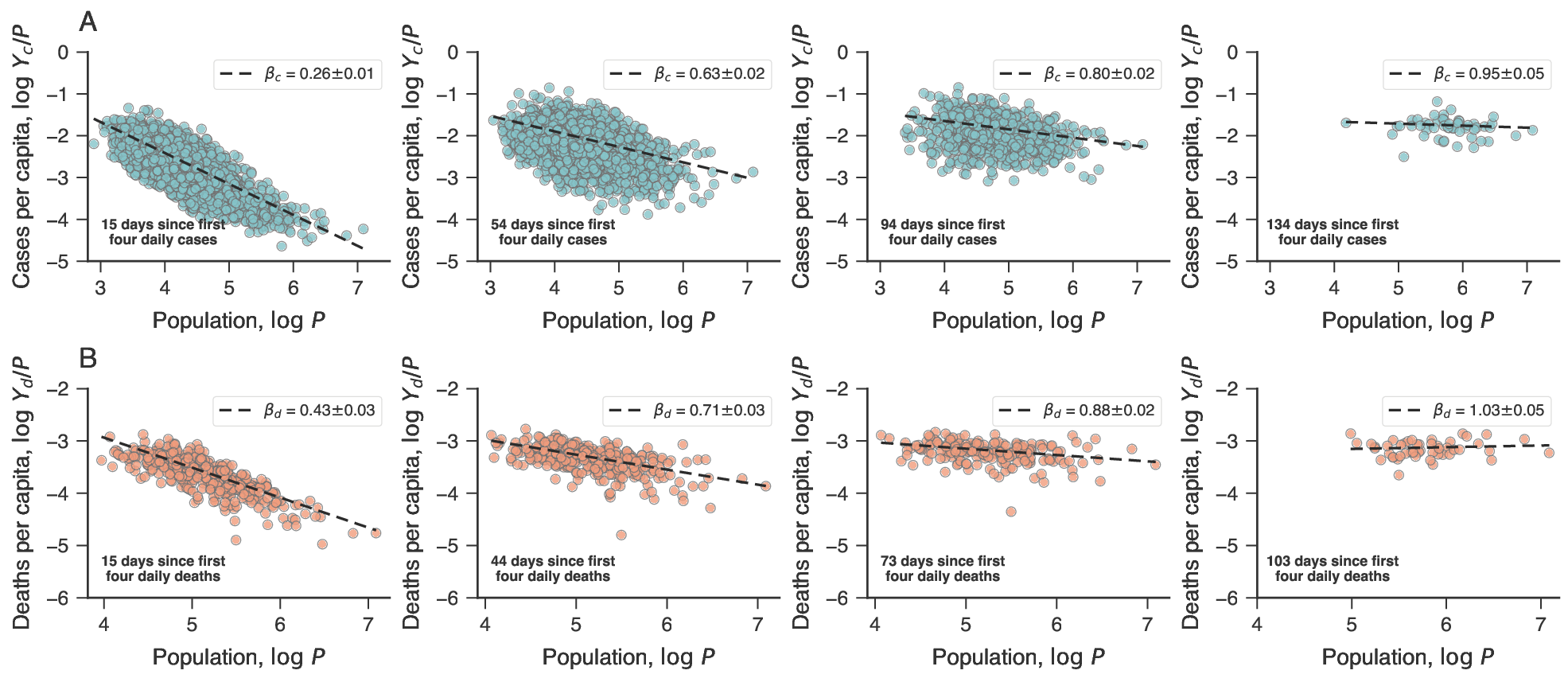}
    \caption{\textbf{Urban scaling relations of COVID-19 cases and deaths \textit{per capita}.} The same as Figure~8 in this Appendix, but considering the first four daily cases and the first four daily deaths as reference points.}
    \label{sfig:fig1_4_per}
\end{figure*}
\clearpage

\begin{figure*}
    \centering
    \includegraphics[width=1\textwidth]{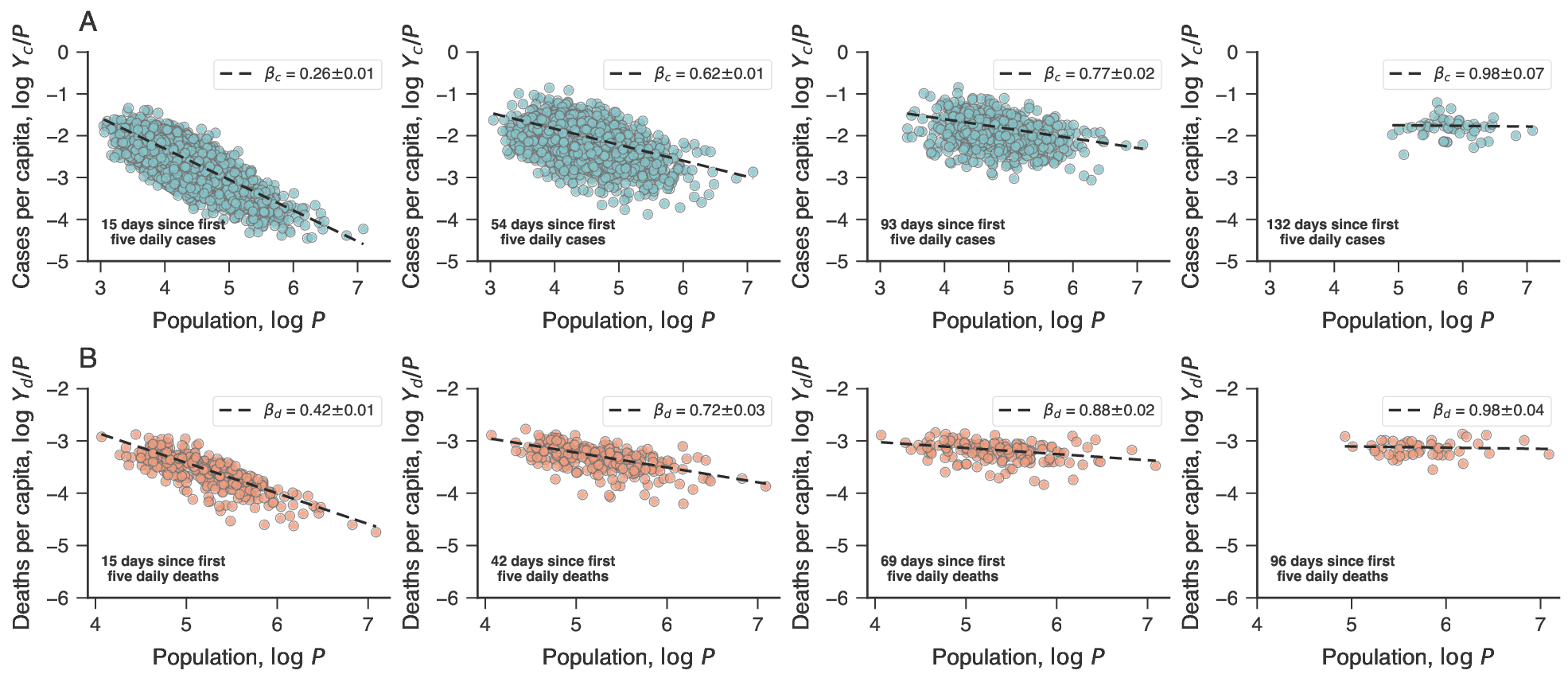}
    \caption{\textbf{Urban scaling relations of COVID-19 cases and deaths \textit{per capita}.} The same as Figure~8 in this Appendix, but considering the first five cases and the first five daily deaths as reference points.}
    \label{sfig:fig1_5_per}
\end{figure*}
\clearpage

\begin{figure*}
    \centering
    \includegraphics[width=1\textwidth]{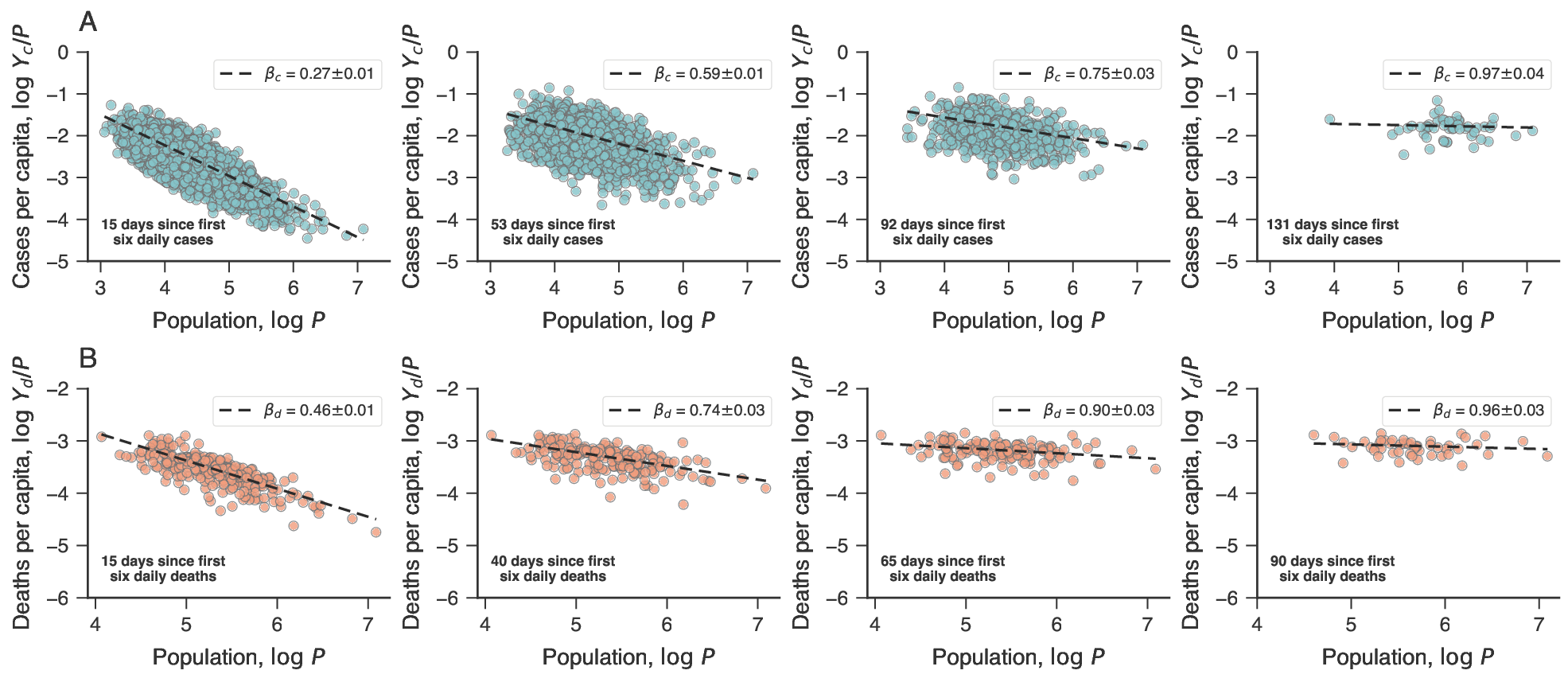}
    \caption{\textbf{Urban scaling relations of COVID-19 cases and deaths \textit{per capita}.} The same as Figure~8 in this Appendix, but considering the first six daily cases and the first six daily deaths as reference points.}
    \label{sfig:fig1_6_per}
\end{figure*}
\clearpage

\begin{figure*}
    \centering
    \includegraphics[width=1\textwidth]{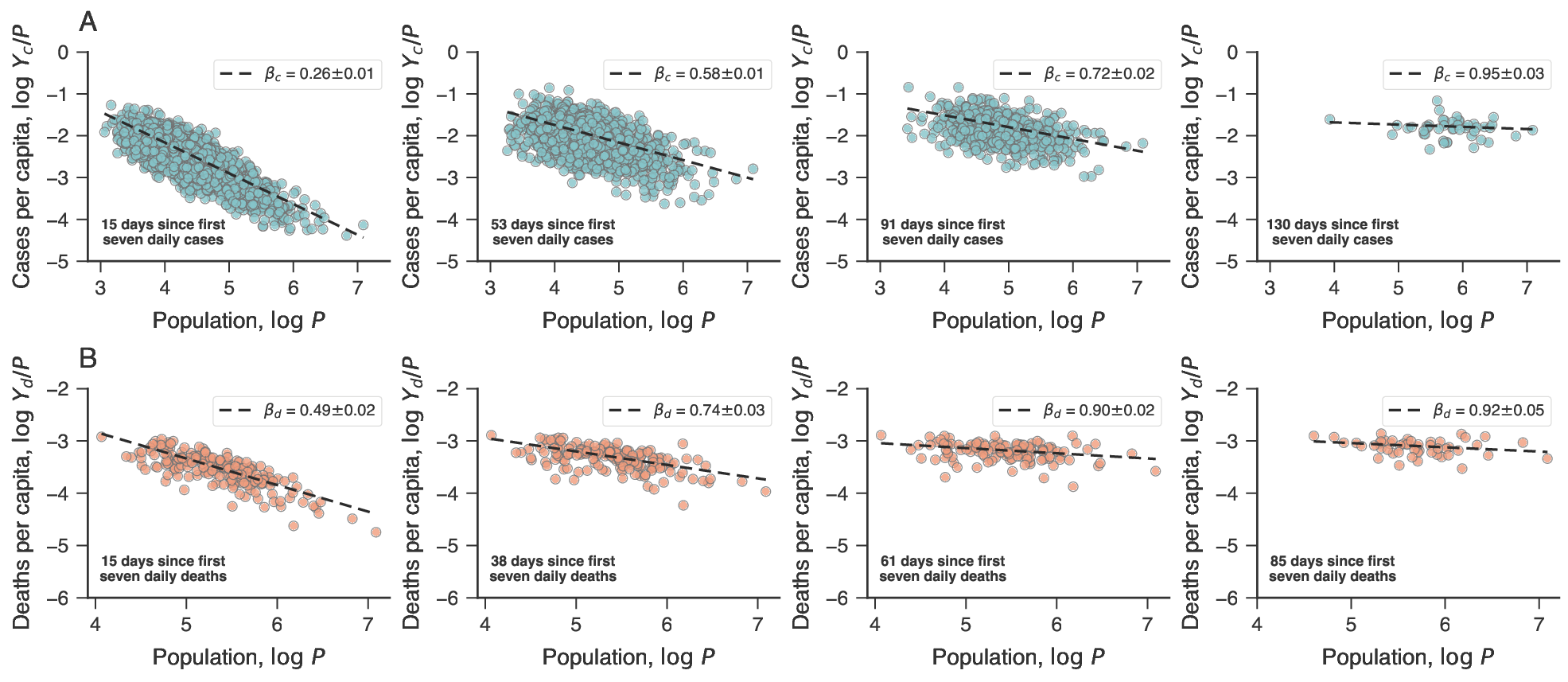}
    \caption{\textbf{Urban scaling relations of COVID-19 cases and deaths \textit{per capita}.} The same as Figure~8 in this Appendix, but considering the first seven daily cases and the first seven daily deaths as reference points.}
    \label{sfig:fig1_7_per}
\end{figure*}
\clearpage

\begin{figure*}
    \centering
    \includegraphics[width=\textwidth]{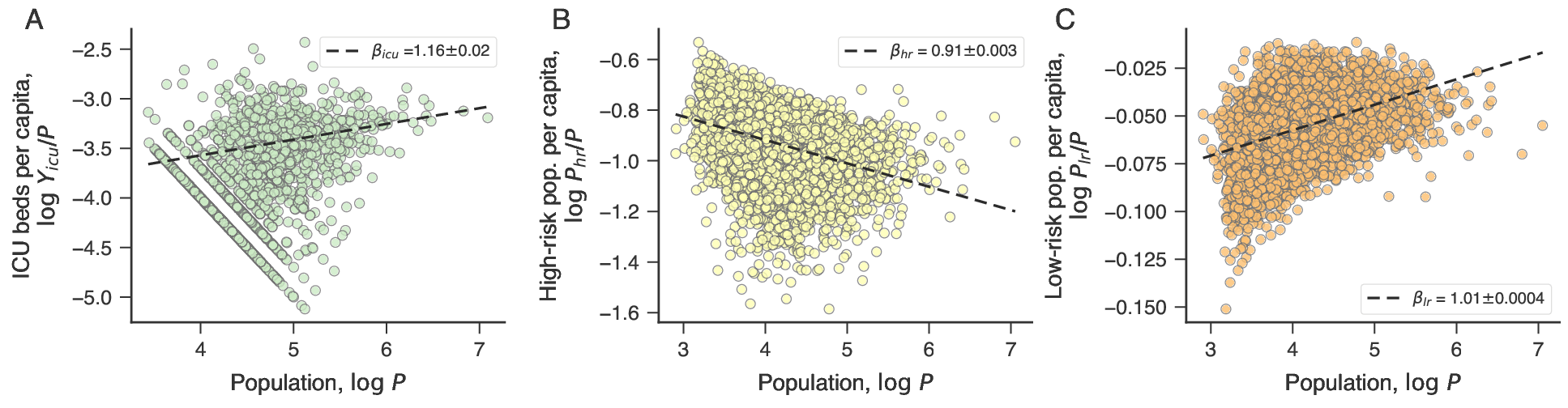}
    \caption{\textbf{Urban scaling of ICU beds, high-risk and low-risk populations \textit{per capita}.} (A) Relationship between the number of ICU beds \textit{per capita} ($Y_{icu}/P$) and the city population ($P$) on logarithmic scale. (B) Relationship between the high-risk population \textit{per capita} ($P_{hr}/P$) and the city population ($P$) on logarithmic scale. (C) Relationship between low-risk population \textit{per capita} ($P_{lr}/P$) and the city population ($P$) on logarithmic scale. In all panels, the dashed lines are the scaling relations with exponents indicated in each plot ($\beta_{icu}-1$ for ICU beds \textit{per capita}, $\beta_{hr}-1$ for high-risk population \textit{per capita}, and $\beta_{lr}-1$ for low-risk population \textit{per capita}).}
    \label{sfig:fig_3}
\end{figure*}

\clearpage

\begin{figure*}
    \centering
    \includegraphics[width=\textwidth]{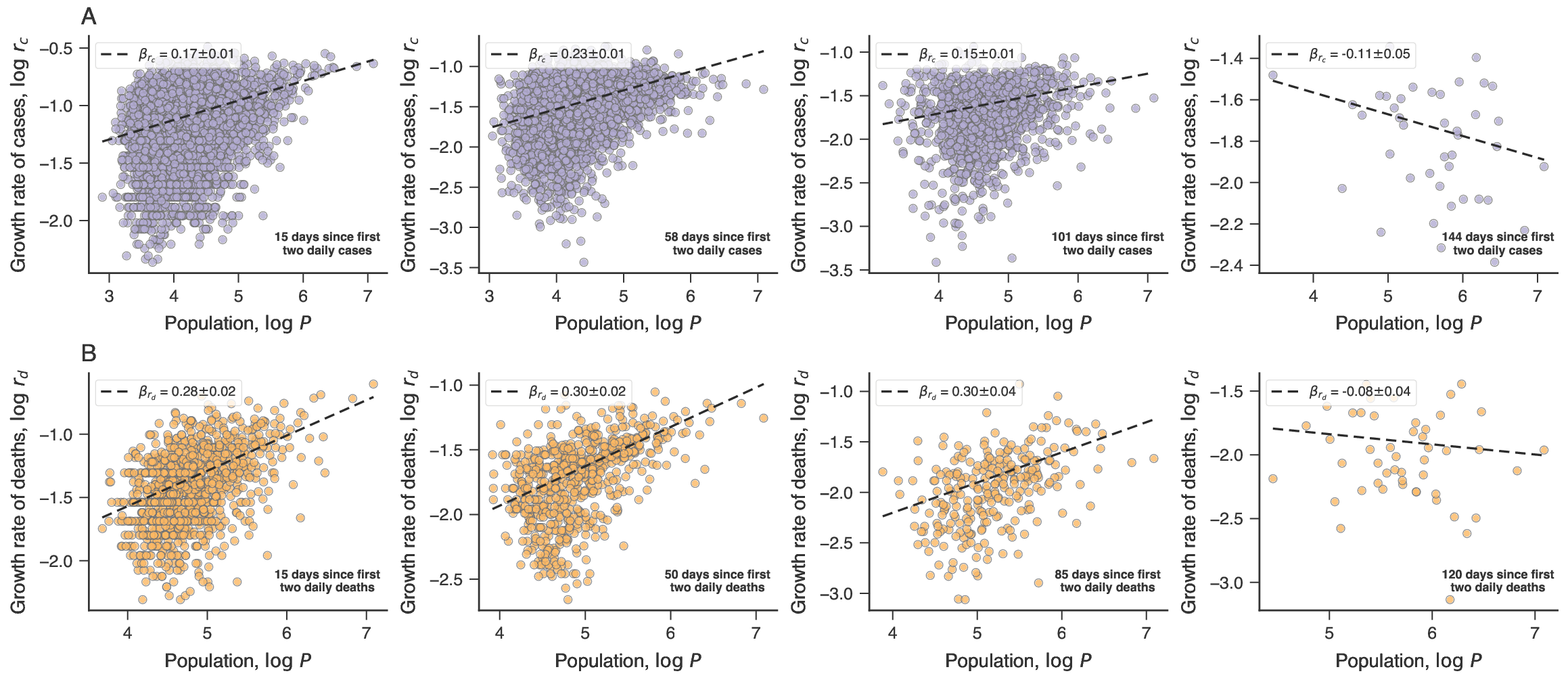}
    \caption{\textbf{Urban scaling relations of growth rates of COVID-19 cases and deaths.} (A) Relationship between the growth rate of cases ($r_c$) and the city population ($P$) on logarithmic scale. Panels show scaling relations for values of $r_c$ estimated after a given number of days since the first two daily cases (four evenly spaced values of $t_c$ between 15 days and the largest value yielding at least 50 cities, as indicated within panels). (B) Relationship between the growth rate of deaths ($r_d$) and the city population ($P$) on logarithmic scale. Panels show scaling relations for values of $r_d$ estimated after a given number of days since the first two daily deaths (four evenly spaced values of $t_d$ between 15 days and the largest value yielding at least 50 cities, as indicated within panels). The markers in (A) and (B) represent cities, and the dashed lines are the adjusted scaling relations with best-fitting exponents indicated in each plot ($\beta_{r_c}$ for the growth rate of cases and $\beta_{r_d}$ for the growth rate of deaths). All rates were estimated using $\tau=14$ as defined in Eq.~(5) of the main text.}
    \label{sfig:growth_rates_2}
\end{figure*}
\clearpage

\begin{figure*}
    \centering
    \includegraphics[width=\textwidth]{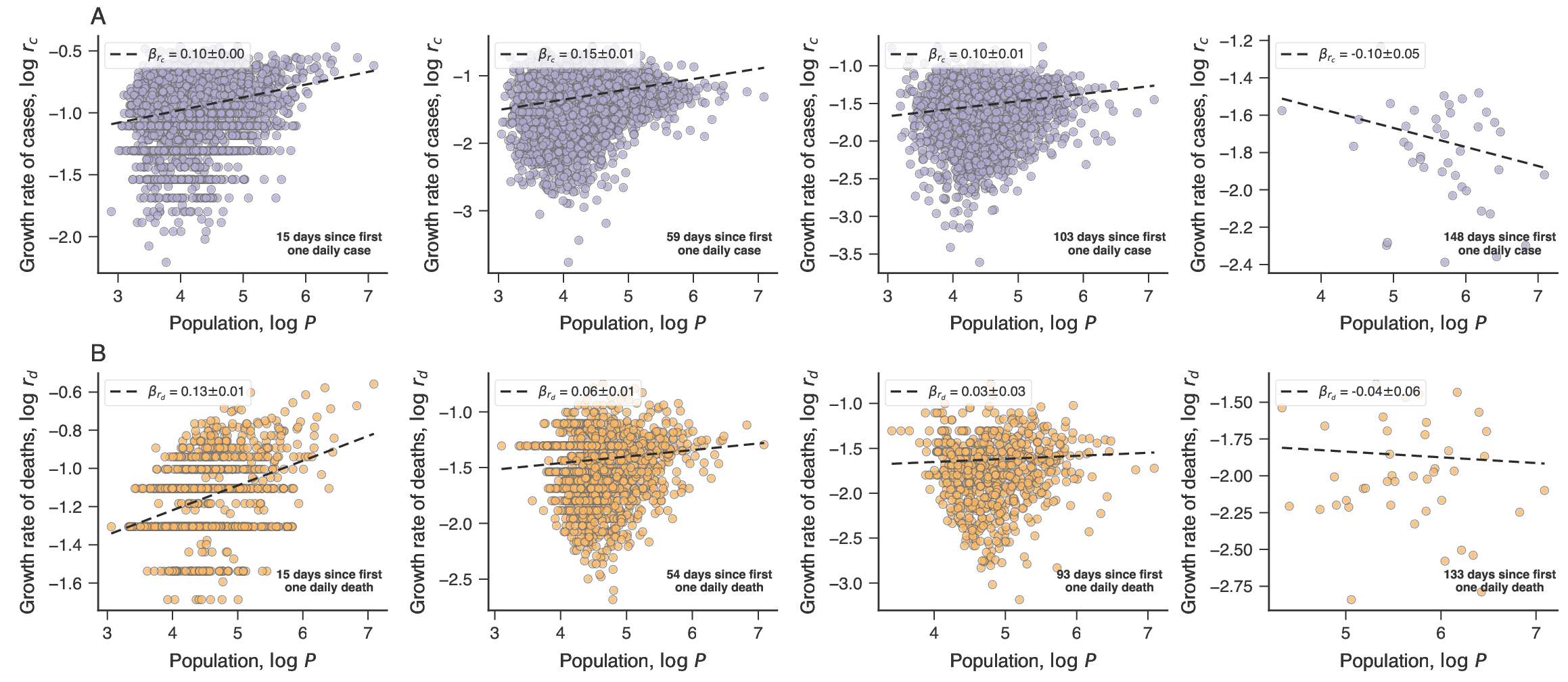}
    \caption{\textbf{Urban scaling relations of growth rates of COVID-19 cases and deaths.} The same as Figure~16 in this Appendix, but considering the first one daily case and first one daily death as reference points.}
    \label{sfig:growth_rates_1}
\end{figure*}
\clearpage

\begin{figure*}
    \centering
    \includegraphics[width=\textwidth]{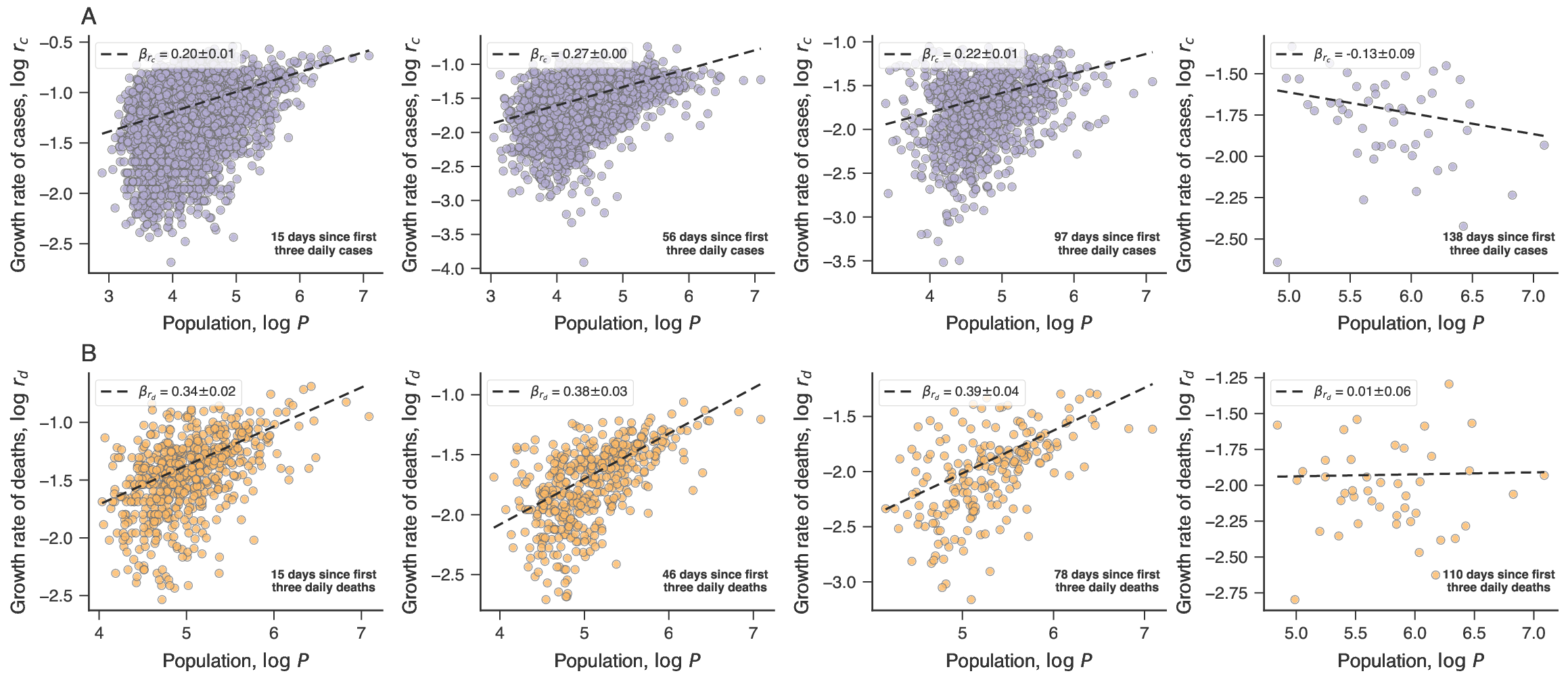}
    \caption{\textbf{Urban scaling relations of growth rates of COVID-19 cases and deaths.} The same as Figure~16 in this Appendix, but considering three daily cases and three daily deaths as reference points.}
    \label{sfig:growth_rates_3}
\end{figure*}
\clearpage

\begin{figure*}
    \centering
    \includegraphics[width=\textwidth]{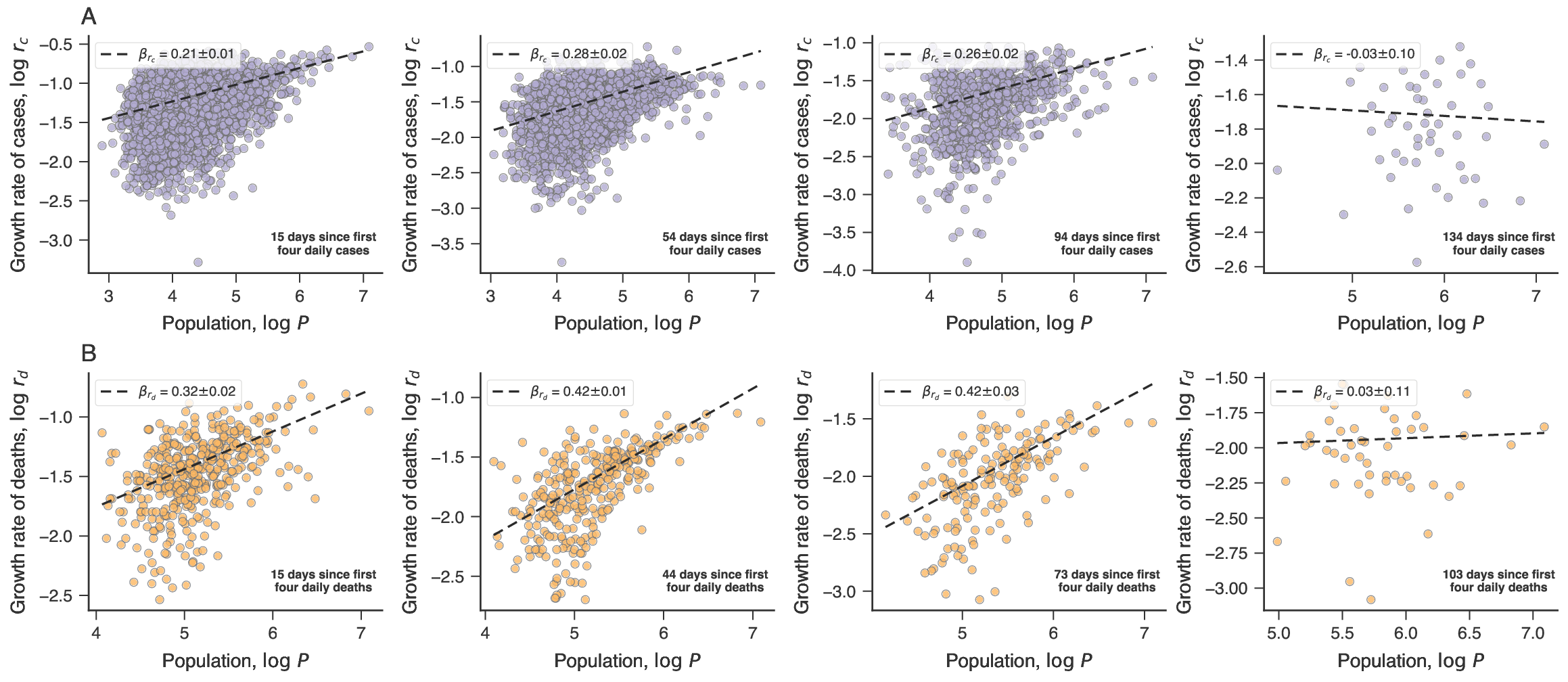}
    \caption{\textbf{Urban scaling relations of growth rates of COVID-19 cases and deaths.} The same as Figure~16 in this Appendix, but considering four daily cases and four daily deaths as reference points.}
    \label{sfig:growth_rates_4}
\end{figure*}
\clearpage

\begin{figure*}
    \centering
    \includegraphics[width=\textwidth]{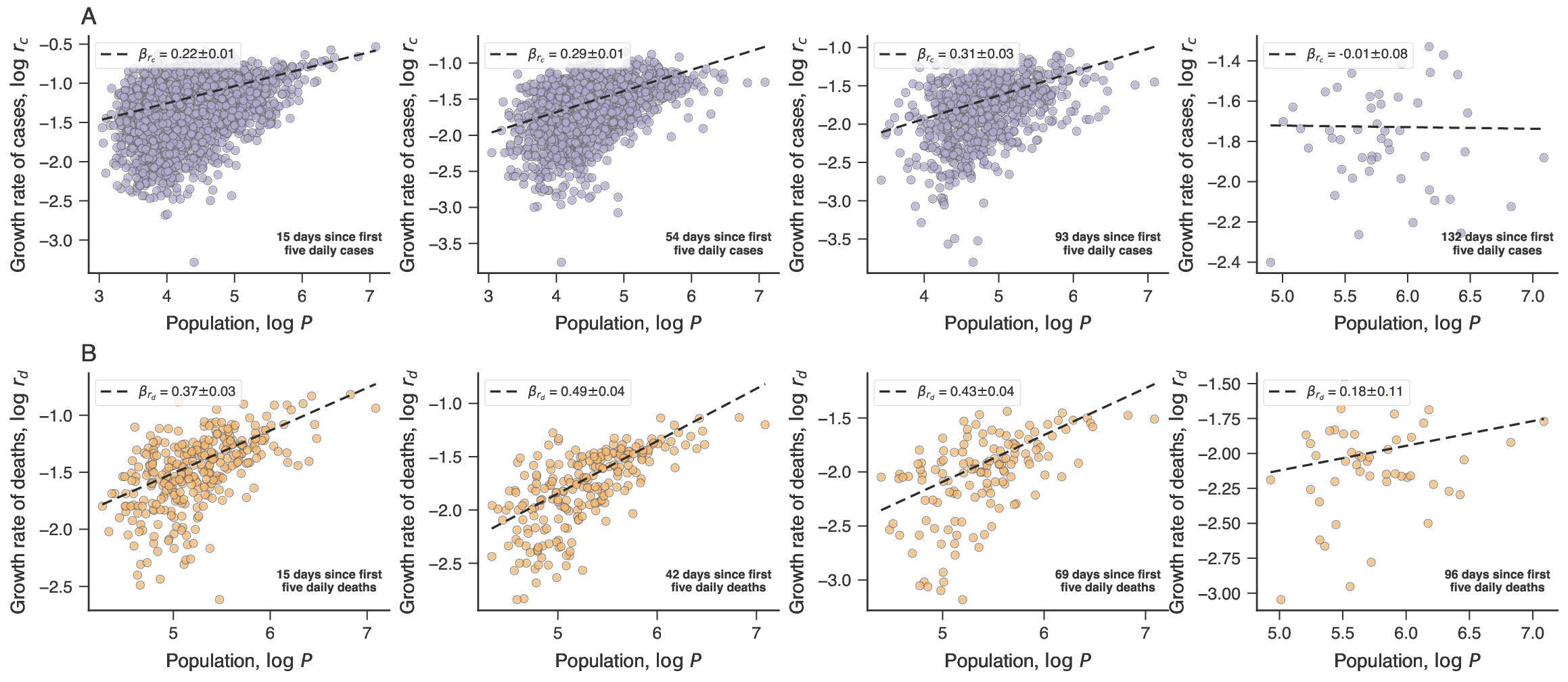}
    \caption{\textbf{Urban scaling relations of growth rates of COVID-19 cases and deaths.} The same as Figure~16 in this Appendix, but considering five daily cases and five daily deaths as reference points.}
    \label{sfig:growth_rates_5}
\end{figure*}
\clearpage

\begin{figure*}
    \centering
    \includegraphics[width=\textwidth]{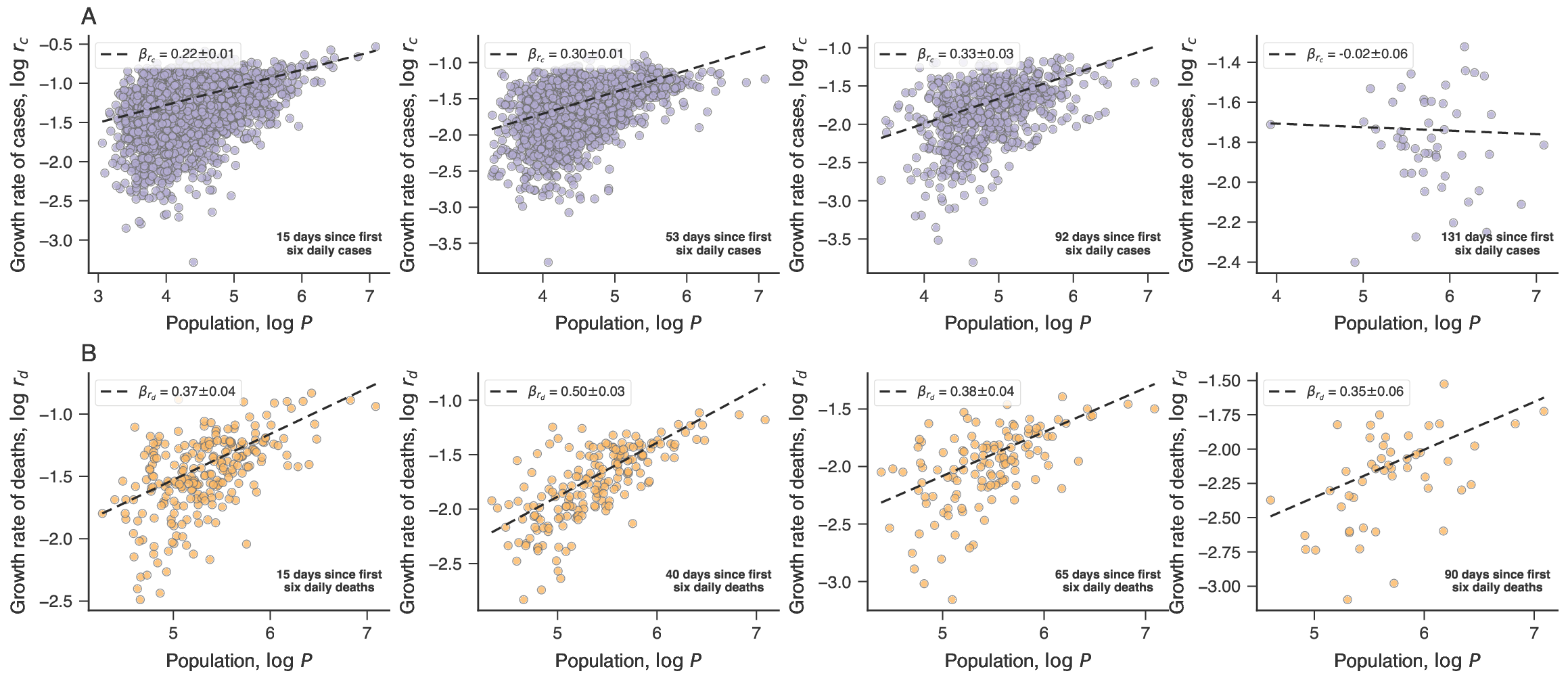}
    \caption{\textbf{Urban scaling relations of growth rates of COVID-19 cases and deaths.} The same as Figure~16 in this Appendix, but considering six daily cases and six daily deaths as reference points.}
    \label{sfig:growth_rates_6}
\end{figure*}
\clearpage

\begin{figure*}
    \centering
    \includegraphics[width=\textwidth]{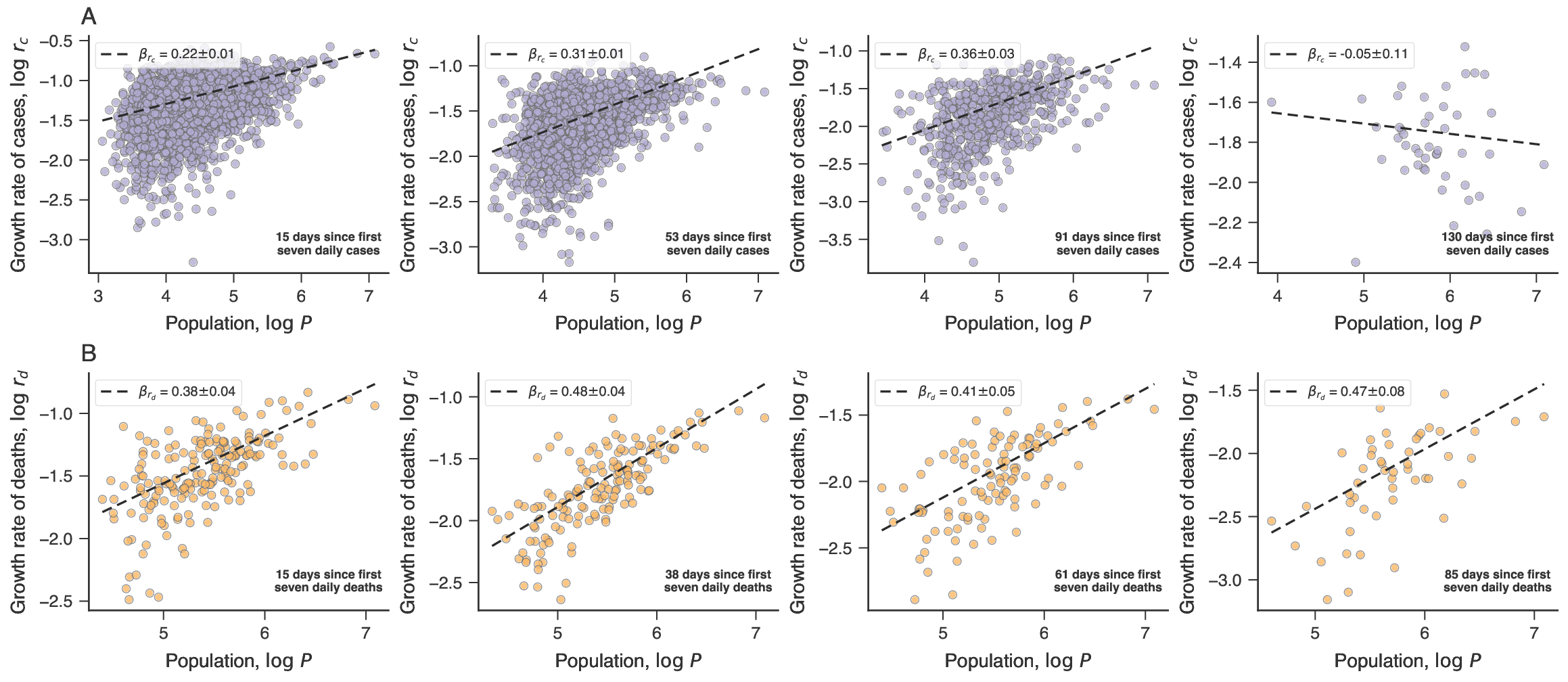}
    \caption{\textbf{Urban scaling relations of growth rates of COVID-19 cases and deaths.} The same as Figure~16 in this Appendix, but considering seven daily cases and seven daily deaths as reference points.}
    \label{sfig:growth_rates_7}
\end{figure*}
\clearpage

\begin{figure*}
    \centering
    \includegraphics[width=0.44\textwidth]{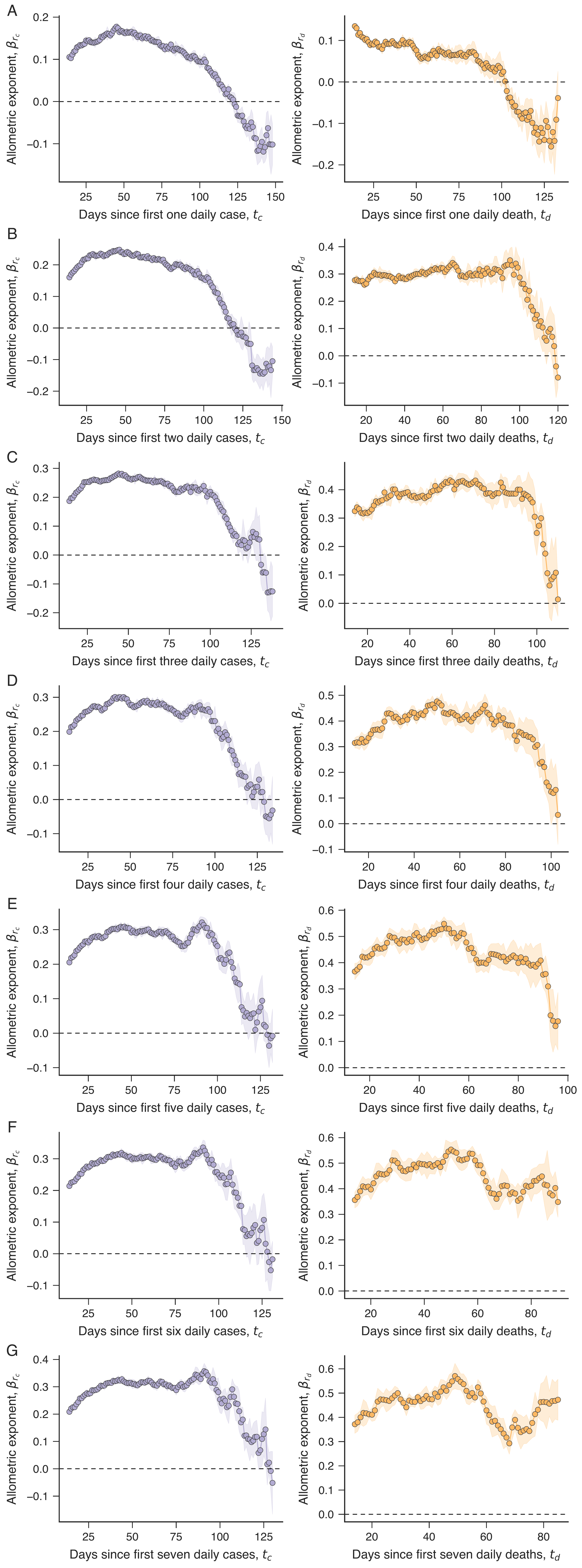}
    \caption{\textbf{Time dependence of the scaling exponents for growth rates of cases and deaths under different choices for the number of daily cases or daily deaths as reference points.} Panels (A)-(G) show the dependence of the exponents $\beta_{r_c}$ and $\beta_{r_d}$ on $t_c$ and $t_d$ when considering the first 1-7 daily cases and the first 1-7 daily deaths as reference points. The shaded regions stand for standard errors, and the horizontal dashed lines represent $\beta_{r_c}=\beta_{r_d}=0$. We notice that behavior observed for large numbers of the reference points appear to follow the behavior of small ones in the long-term course of the pandemic.}
    \label{sfig:exp_growth_rates}
\end{figure*}
\clearpage

\begin{figure*}
    \centering
    \includegraphics[width=0.4\textwidth]{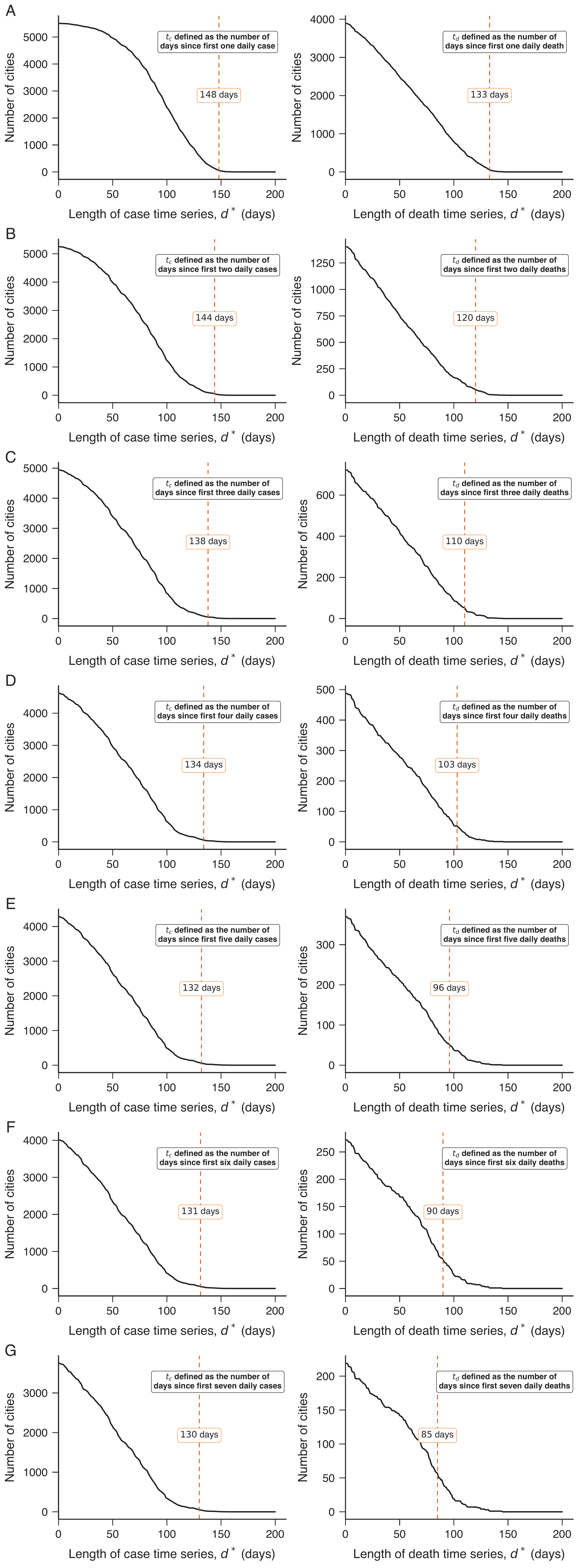}
    \caption{\textbf{Number of cities with time series longer than a particular number of days.} Left panels show the number of cities reporting cases of COVID-19 with time series larger than $d^*$ days. The vertical dashed lines indicate that there are 50 cities with confirmed cases time series longer than the particular of number of days indicated within the plots. Right panels show the number of cities reporting deaths caused by COVID-19 with time series longer than $d^*$ days. The vertical dashed lines indicate that there are 50 cities with deaths time series longer than the particular of number of days indicated within the plots. We have used these thresholds to ensure that our scaling relations are estimated from samples sizes having at least 50 cities. Panels (A)-(G) show the results when considering the first 1-7 daily cases and the first 1-7 daily deaths as reference points.}
    \label{sfig:len_ts}
\end{figure*}
\clearpage

\begin{figure*}
    \centering
    \includegraphics[width=0.7\textwidth]{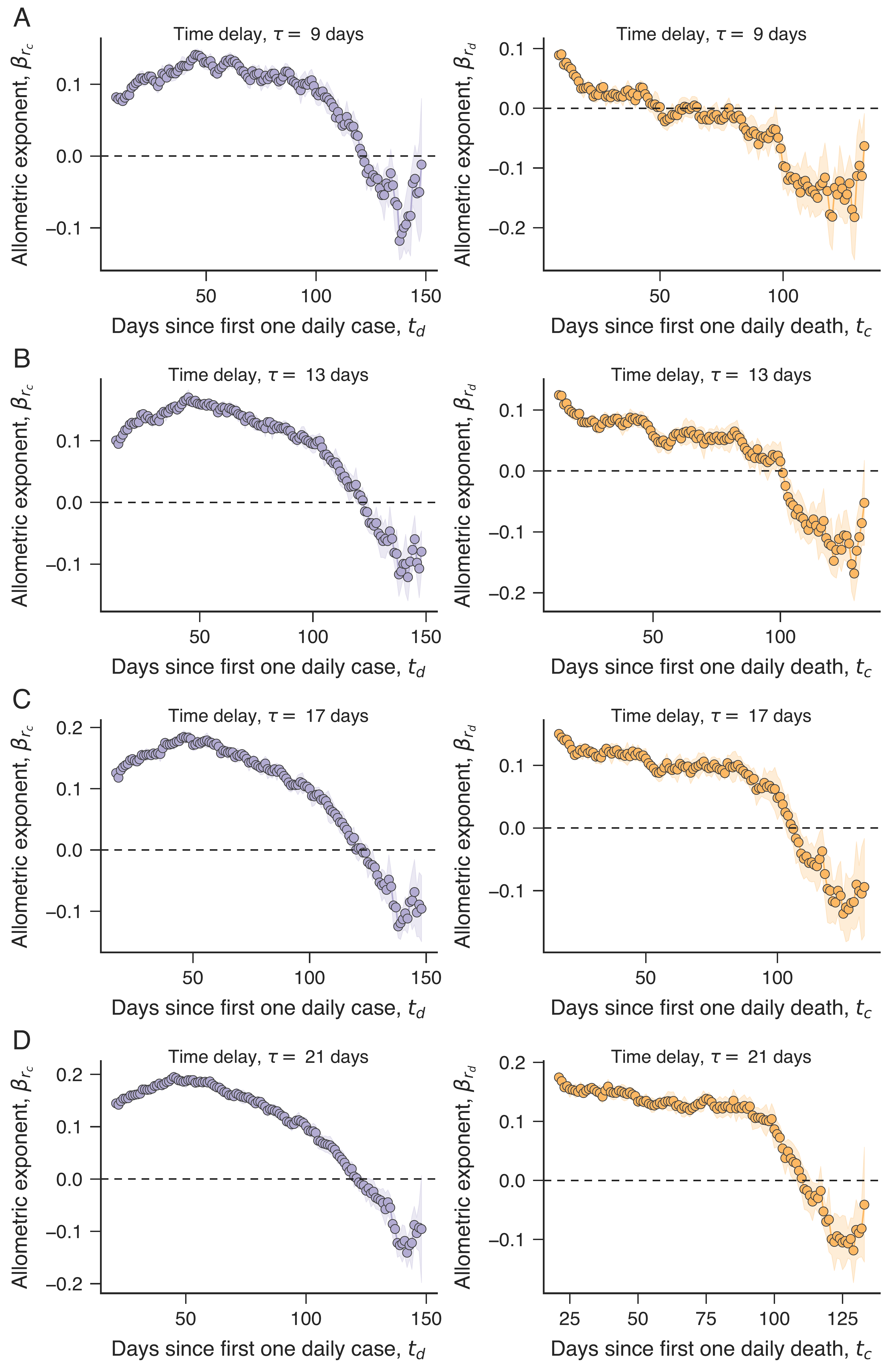}
    \caption{\textbf{Variations in the scaling exponents for the growth rates of cases and deaths under different values of time delay $\tau$.} Panels (A)-(D) show the dependence of the exponent $\beta_{r_c}$ (left panels) and $\beta_{r_d}$ (right panels) on the number of days since the first one daily case ($t_c$) or death ($t_d$) for different values of $\tau$.}
    \label{sfig:different_rates_1}
\end{figure*}
\clearpage

\begin{figure*}
    \centering
    \includegraphics[width=0.7\textwidth]{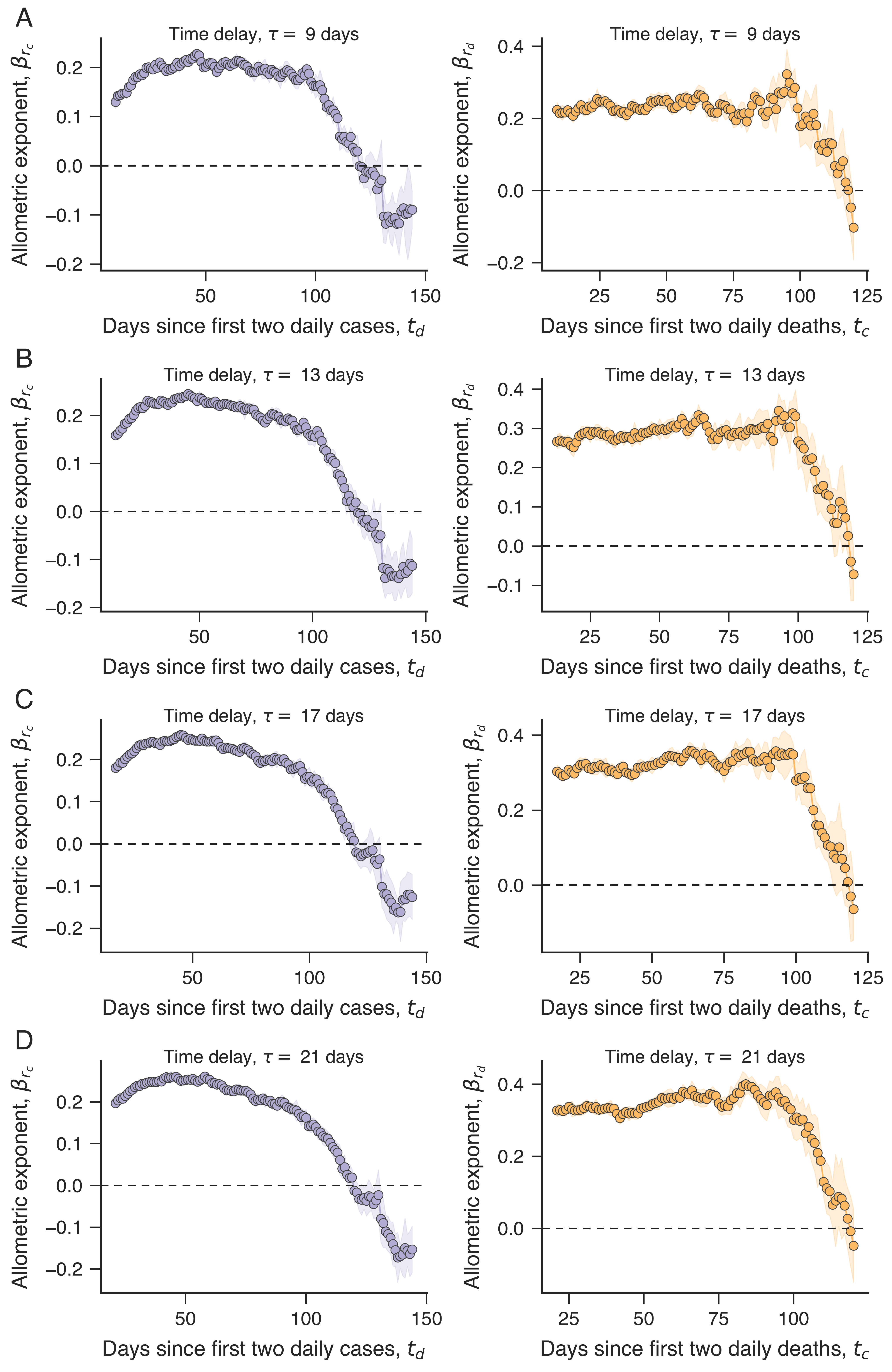}
    \caption{\textbf{Variations in the scaling exponents for the growth rates of cases and deaths under different values of time delay $\tau$.} Panels (A)-(D) show the dependence of the exponent $\beta_{r_c}$ (left panels) and $\beta_{r_d}$ (right panels) on the number of days since the first two daily cases ($t_c$) or deaths ($t_d$) for different values of $\tau$.}
    \label{sfig:different_rates_2}
\end{figure*}
\clearpage

\begin{figure*}
    \centering
    \includegraphics[width=0.7\textwidth]{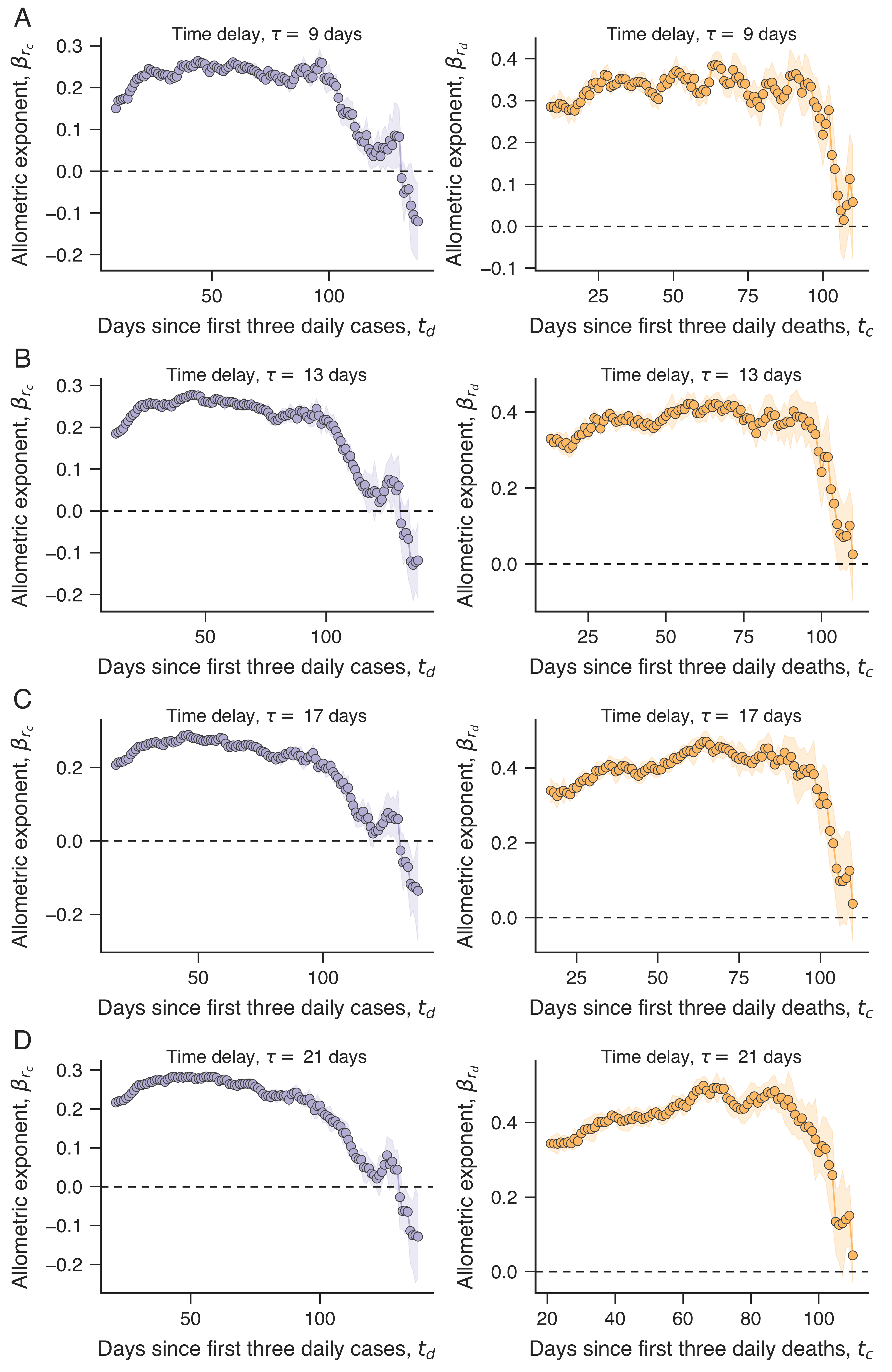}
    \caption{\textbf{Variations in the scaling exponents for the growth rates of cases and deaths under different values of time delay $\tau$.} Panels (A)-(D) show the dependence of the exponent $\beta_{r_c}$ (left panels) and $\beta_{r_d}$ (right panels) on the number of days since the first three daily cases ($t_c$) or deaths ($t_d$) for different values of $\tau$.}
    \label{sfig:different_rates_3}
\end{figure*}
\clearpage

\begin{figure*}
    \centering
    \includegraphics[width=0.7\textwidth]{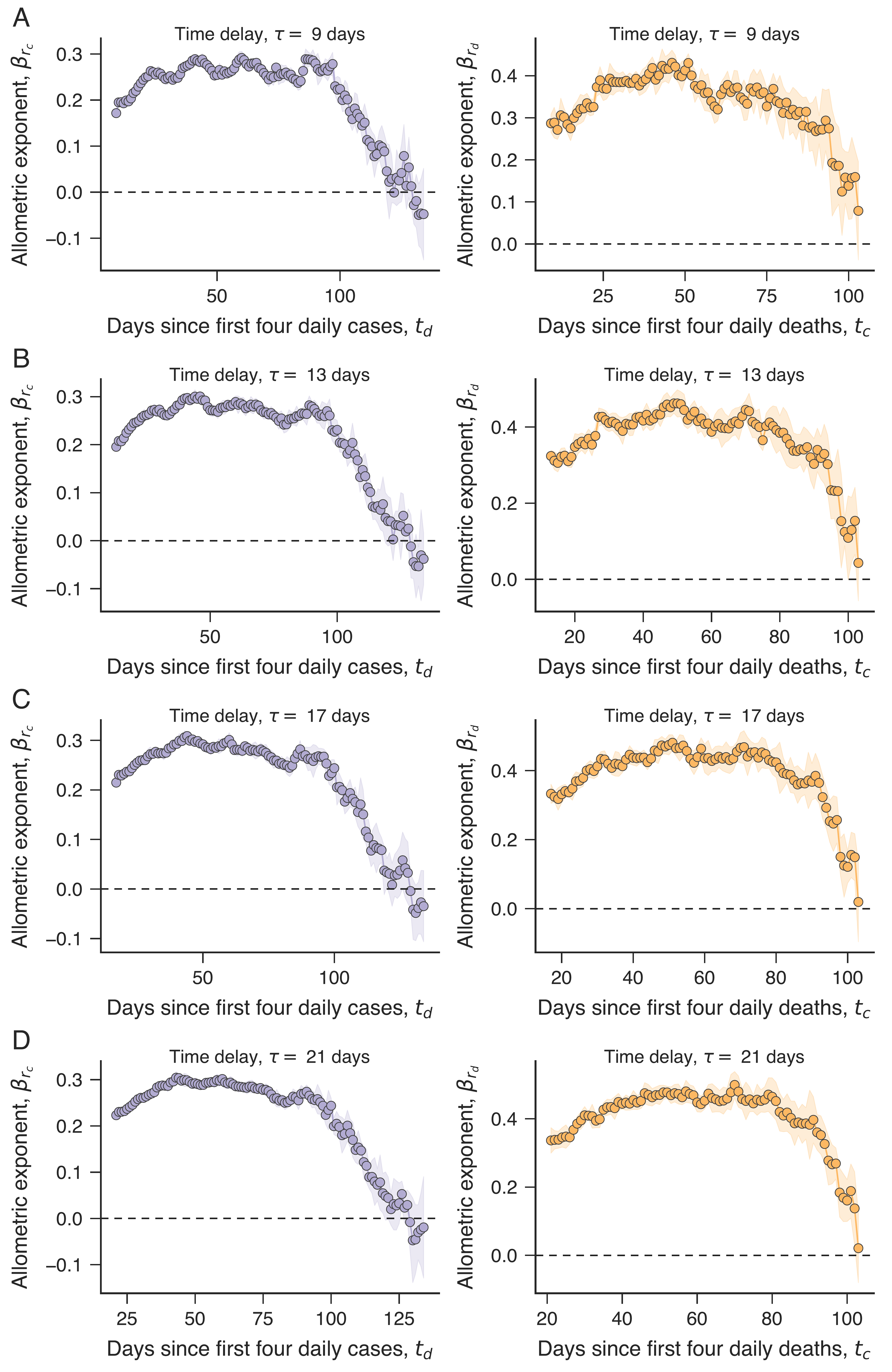}
    \caption{\textbf{Variations in the scaling exponents for the growth rates of cases and deaths under different values of time delay $\tau$.} Panels (A)-(D) show the dependence of the exponent $\beta_{r_c}$ (left panels) and $\beta_{r_d}$ (right panels) on the number of days since the first four daily cases ($t_c$) or deaths ($t_d$) for different values of $\tau$.}
    \label{sfig:different_rates_4}
\end{figure*}
\clearpage

\begin{figure*}
    \centering
    \includegraphics[width=0.7\textwidth]{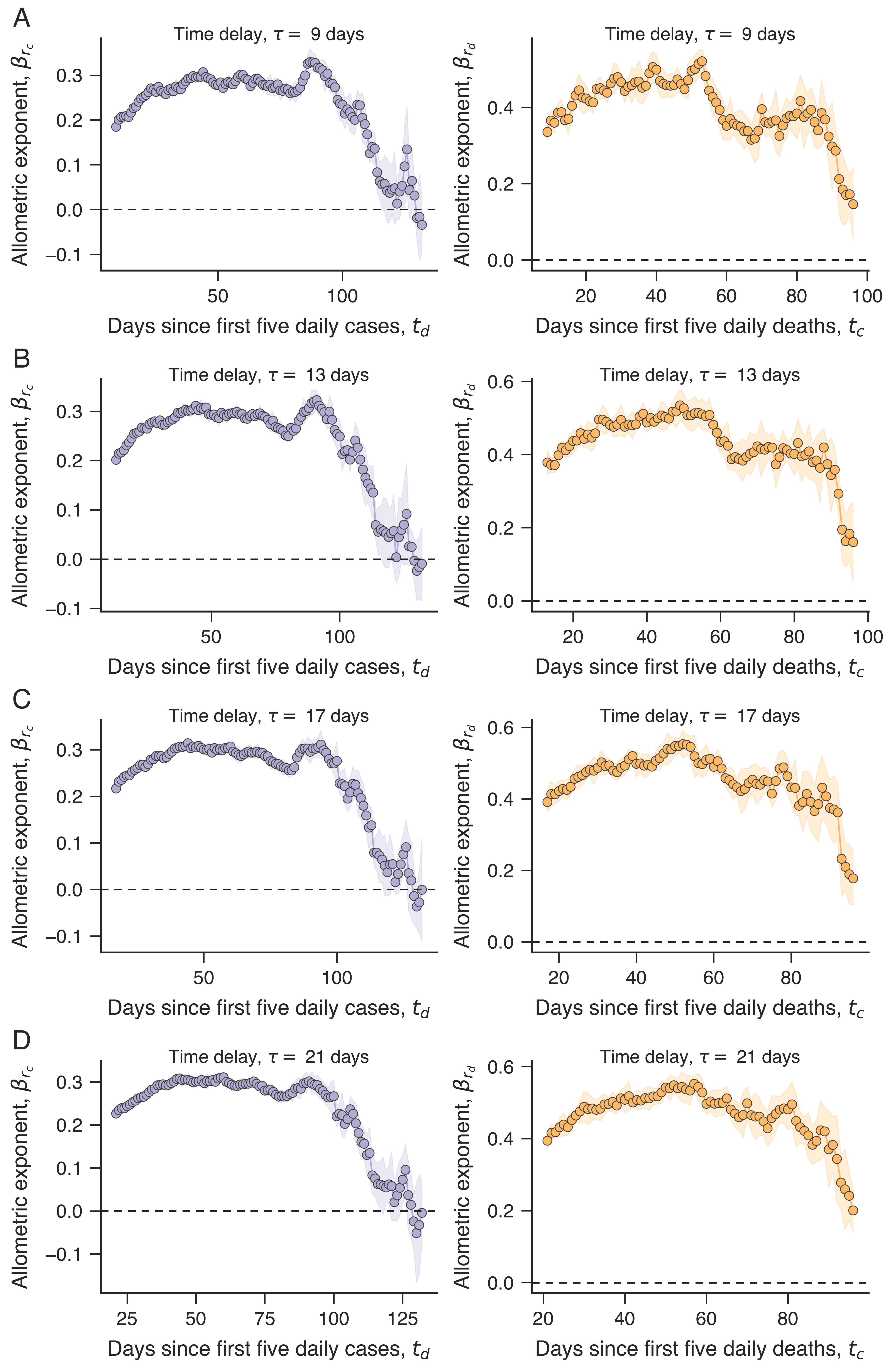}
    \caption{\textbf{Variations in the scaling exponents for the growth rates of cases and deaths under different values of time delay $\tau$.} Panels (A)-(D) show the dependence of the exponent $\beta_{r_c}$ (left panels) and $\beta_{r_d}$ (right panels) on the number of days since the first five daily cases ($t_c$) or deaths ($t_d$) for different values of $\tau$.}
    \label{sfig:different_rates_5}
\end{figure*}
\clearpage

\begin{figure*}
    \centering
    \includegraphics[width=0.7\textwidth]{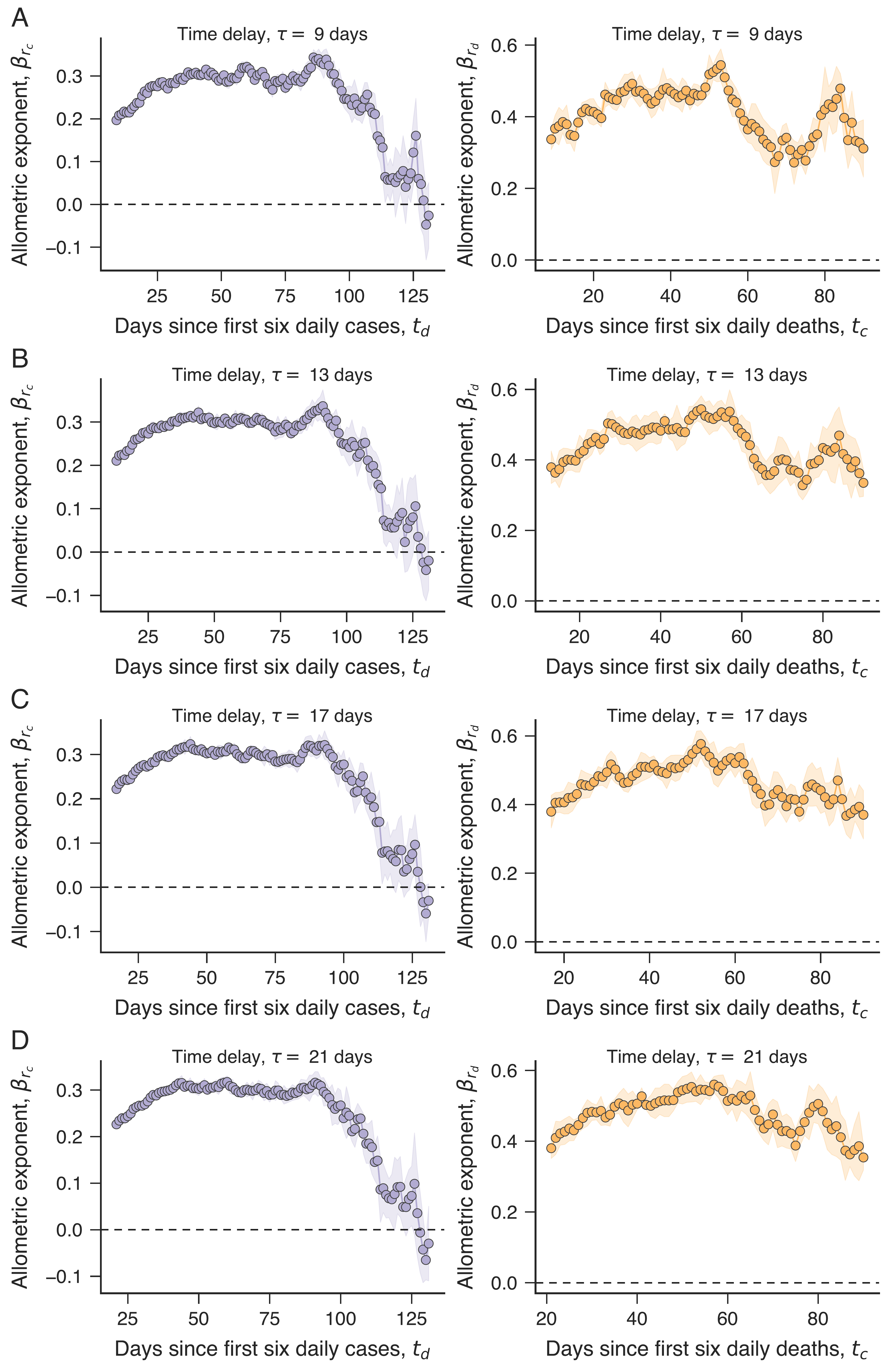}
    \caption{\textbf{Variations in the scaling exponents for the growth rates of cases and deaths under different values of time delay $\tau$.} Panels (A)-(D) show the dependence of the exponent $\beta_{r_c}$ (left panels) and $\beta_{r_d}$ (right panels) on the number of days since the first six daily cases ($t_c$) or deaths ($t_d$) for different values of $\tau$.}
    \label{sfig:different_rates_6}
\end{figure*}
\clearpage

\begin{figure*}
    \centering
    \includegraphics[width=0.7\textwidth]{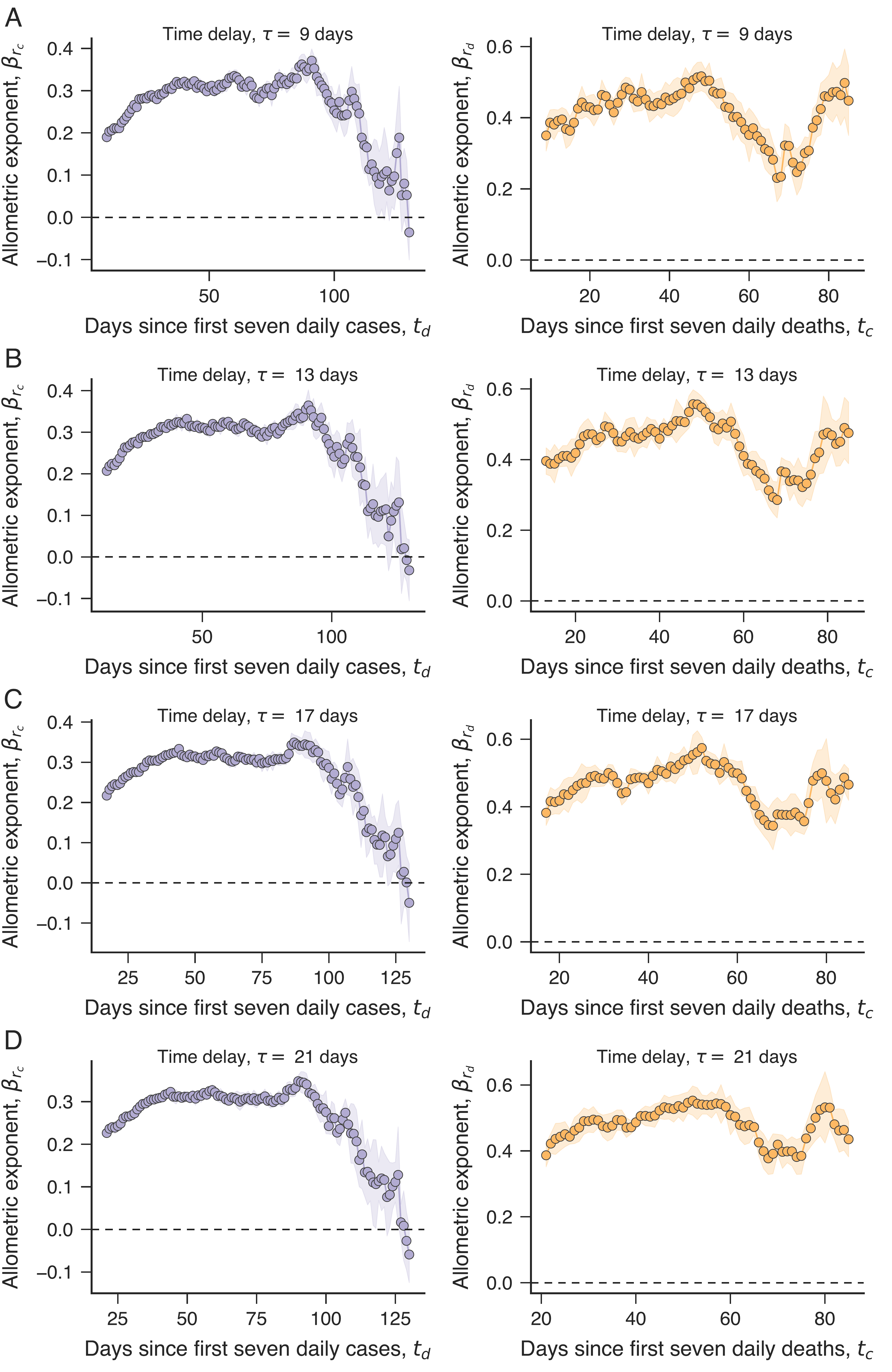}
    \caption{\textbf{Variations in the scaling exponents for the growth rates of cases and deaths under different values of time delay $\tau$.} Panels (A)-(D) show the dependence of the exponent $\beta_{r_c}$ (left panels) and $\beta_{r_d}$ (right panels) on the number of days since the first seven daily cases ($t_c$) or deaths ($t_d$) for different values of $\tau$.}
    \label{sfig:different_rates_7}
\end{figure*}
\clearpage

\begin{figure*}
    \centering
    \includegraphics[width=0.8\textwidth]{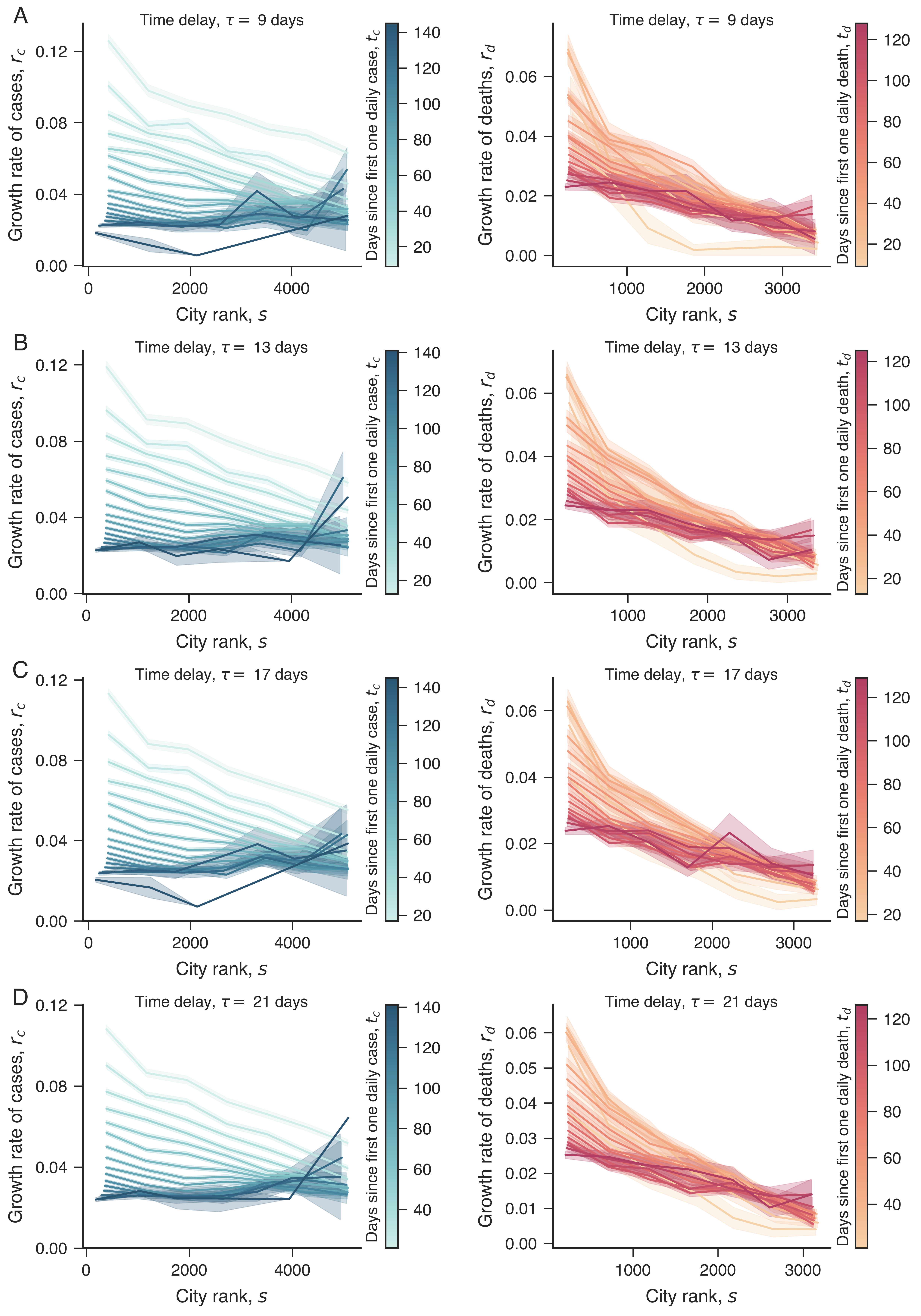}
    \caption{\textbf{Variations in the association between the growth rates and the city rank under different values of time delay $\tau$.} Panels (A)-(D) show the average relationship between the growth rates of COVID-19 cases ($r_c$, left panels) and deaths ($r_d$, right panels) and the city rank $s$ for number of days since the first one daily case ($t_c$) or death ($t_d$) and for different values of $\tau$ (indicated within the plots).}
    \label{sfig:rank_rates_1}
\end{figure*}
\clearpage

\begin{figure*}
    \centering
    \includegraphics[width=0.8\textwidth]{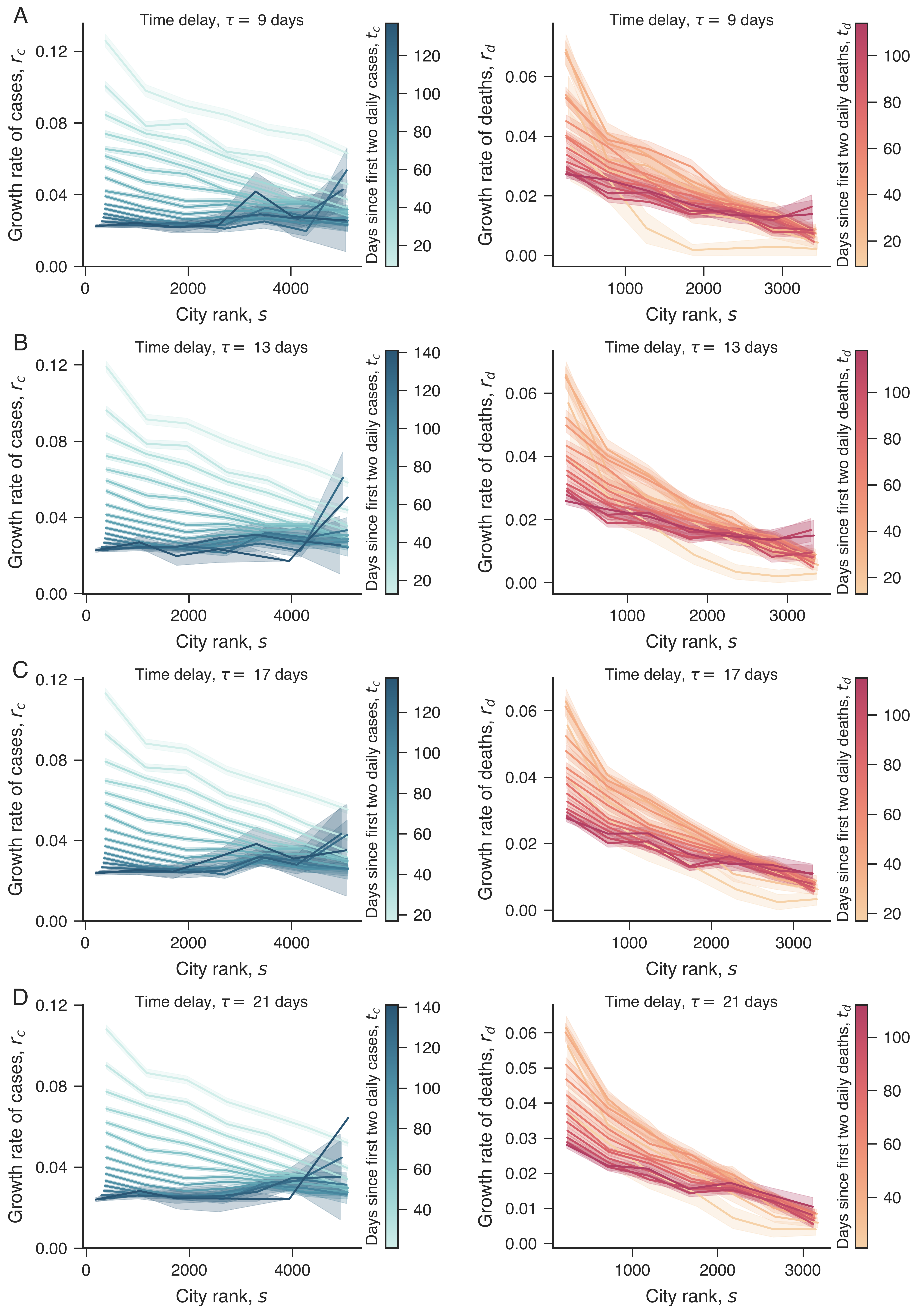}
    \caption{\textbf{Variations in the association between the growth rates and the city rank under different values of time delay $\tau$.} Panels (A)-(D) show the average relationship between the growth rates of COVID-19 cases ($r_c$, left panels) and deaths ($r_d$, right panels) and the city rank $s$ for number of days since the first two daily cases ($t_c$) or deaths ($t_d$) and for different values of $\tau$ (indicated within the plots).}
    \label{sfig:rank_rates_2}
\end{figure*}
\clearpage

\begin{figure*}
    \centering
    \includegraphics[width=0.8\textwidth]{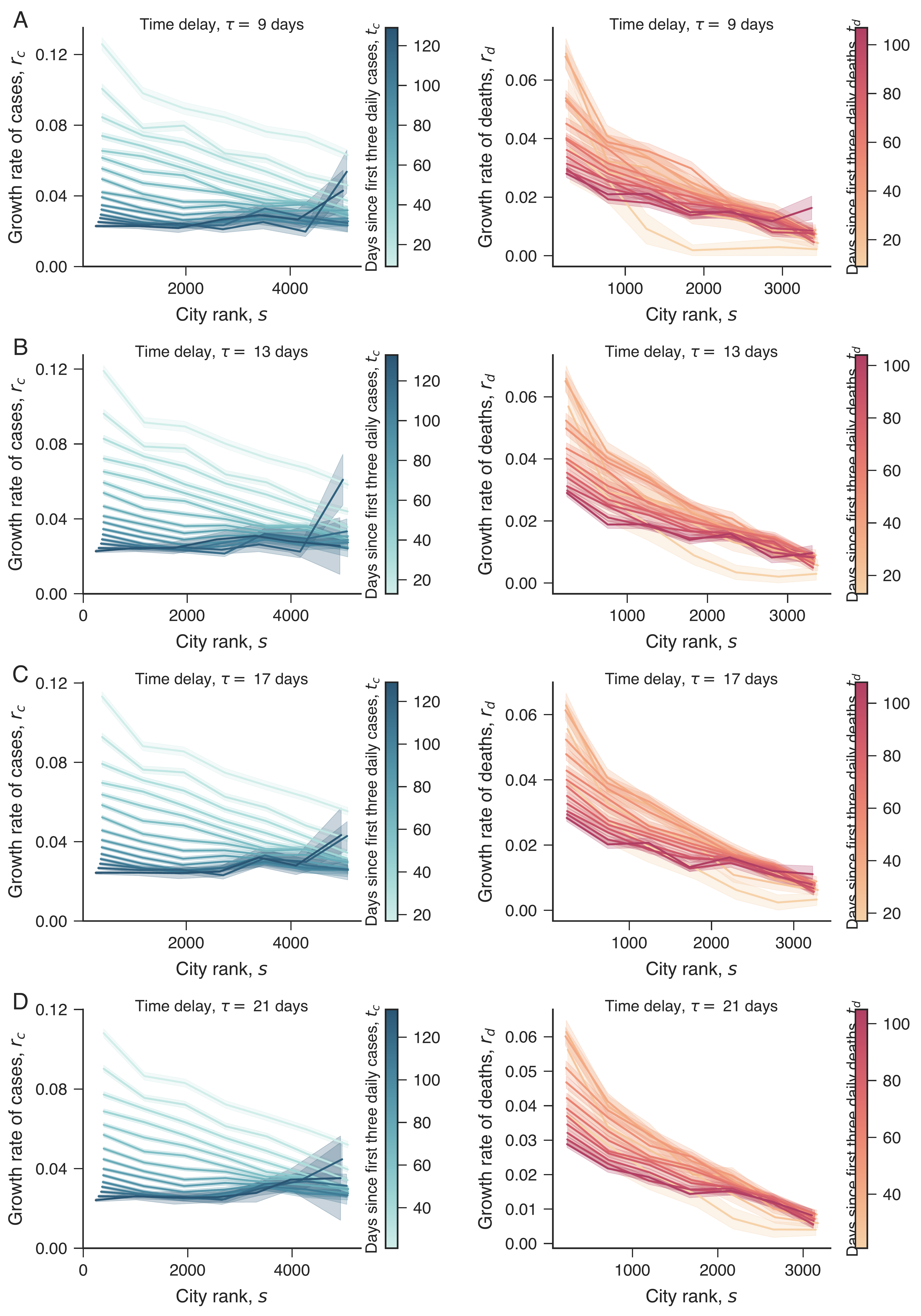}
    \caption{\textbf{Variations in the association between the growth rates and the city rank under different values of time delay $\tau$.} Panels (A)-(D) show the average relationship between the growth rates of COVID-19 cases ($r_c$, left panels) and deaths ($r_d$, right panels) and the city rank $s$ for number of days since the first three daily cases ($t_c$) or deaths ($t_d$) and for different values of $\tau$ (indicated within the plots).}
    \label{sfig:rank_rates_3}
\end{figure*}
\clearpage

\begin{figure*}
    \centering
    \includegraphics[width=0.8\textwidth]{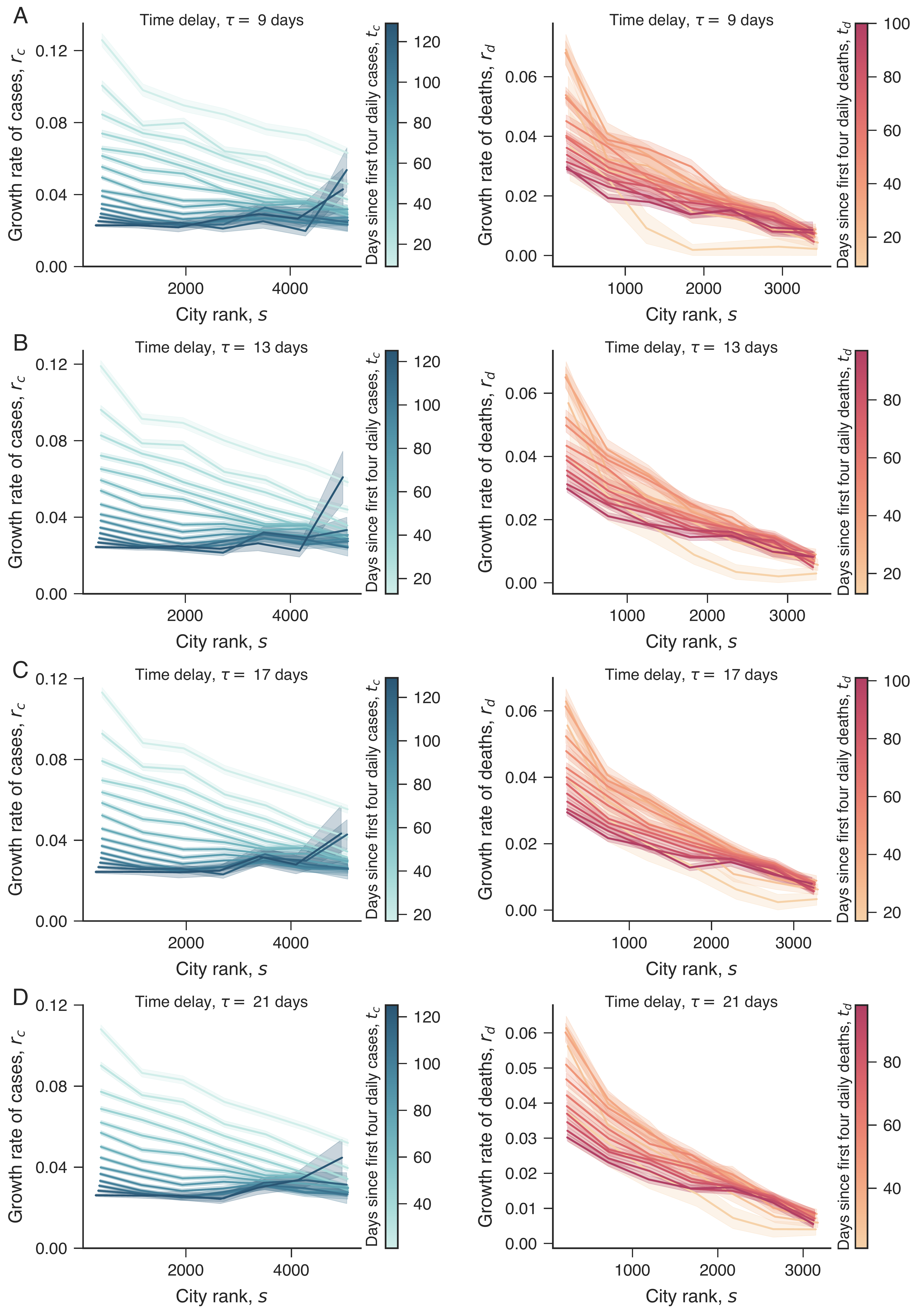}
    \caption{\textbf{Variations in the association between the growth rates and the city rank under different values of time delay $\tau$.} Panels (A)-(D) show the average relationship between the growth rates of COVID-19 cases ($r_c$, left panels) and deaths ($r_d$, right panels) and the city rank $s$ for number of days since the first four daily cases ($t_c$) or deaths ($t_d$) and for different values of $\tau$ (indicated within the plots).}
    \label{sfig:rank_rates_4}
\end{figure*}
\clearpage

\begin{figure*}
    \centering
    \includegraphics[width=0.8\textwidth]{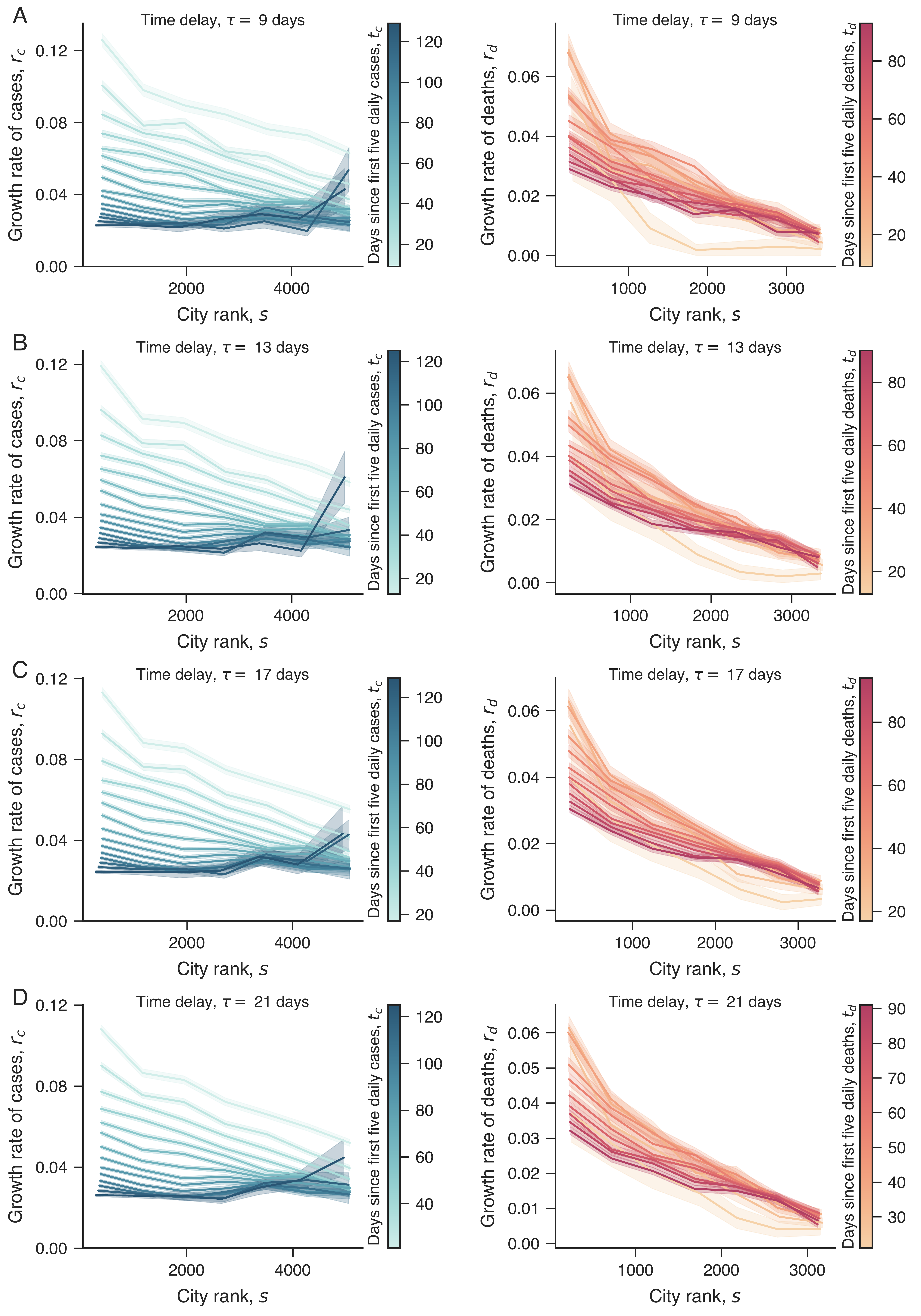}
    \caption{\textbf{Variations in the association between the growth rates and the city rank under different values of time delay $\tau$.} Panels (A)-(D) show the average relationship between the growth rates of COVID-19 cases ($r_c$, left panels) and deaths ($r_d$, right panels) and the city rank $s$ for number of days since the first five daily cases ($t_c$) or deaths ($t_d$) and for different values of $\tau$ (indicated within the plots).}
    \label{sfig:rank_rates_5}
\end{figure*}
\clearpage

\begin{figure*}
    \centering
    \includegraphics[width=0.8\textwidth]{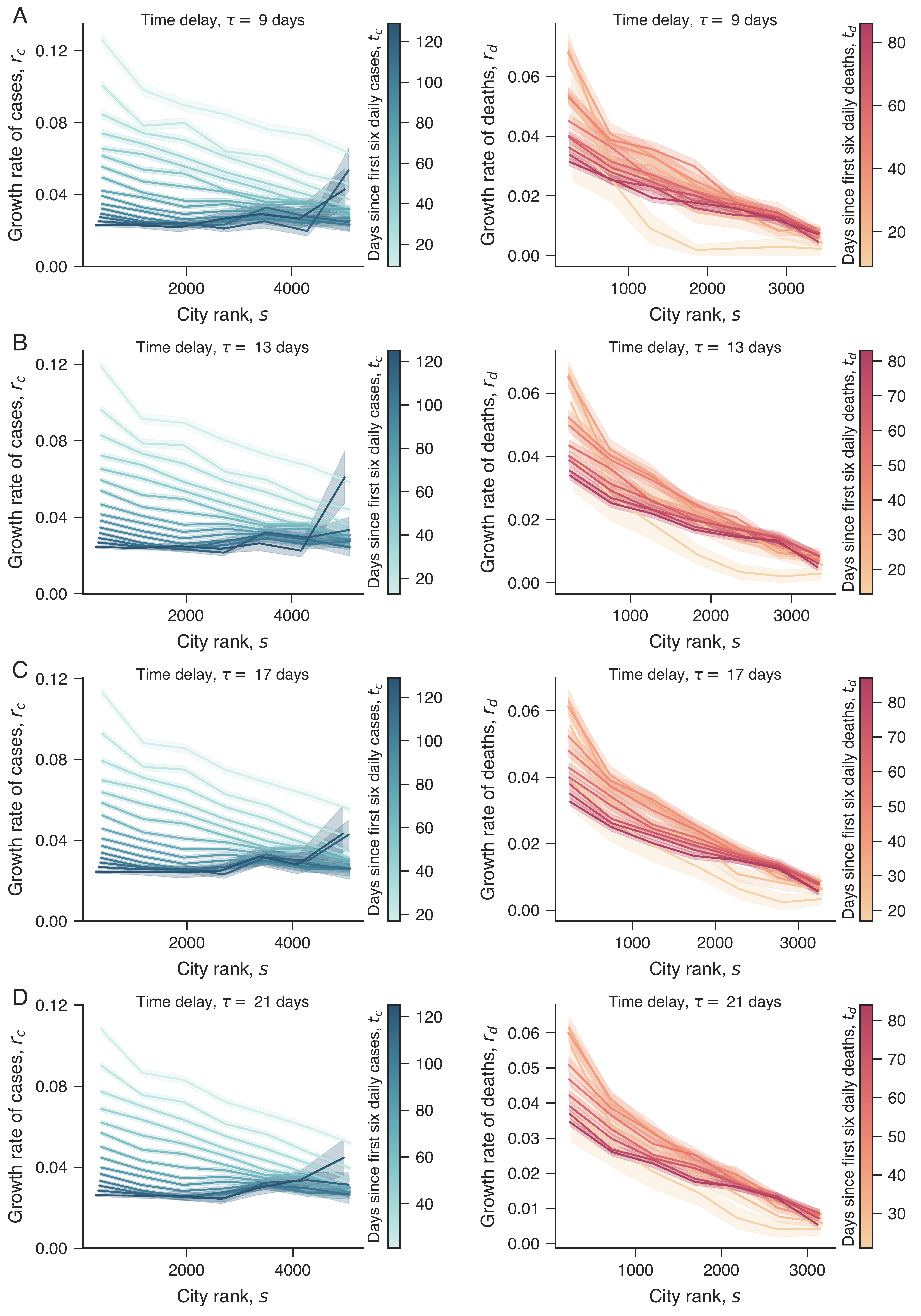}
    \caption{\textbf{Variations in the association between the growth rates and the city rank under different values of time delay $\tau$.} Panels (A)-(D) show the average relationship between the growth rates of COVID-19 cases ($r_c$, left panels) and deaths ($r_d$, right panels) and the city rank $s$ for number of days since the first six daily cases ($t_c$) or deaths ($t_d$) and for different values of $\tau$ (indicated within the plots).}
    \label{sfig:rank_rates_6}
\end{figure*}
\clearpage

\begin{figure*}
    \centering
    \includegraphics[width=0.8\textwidth]{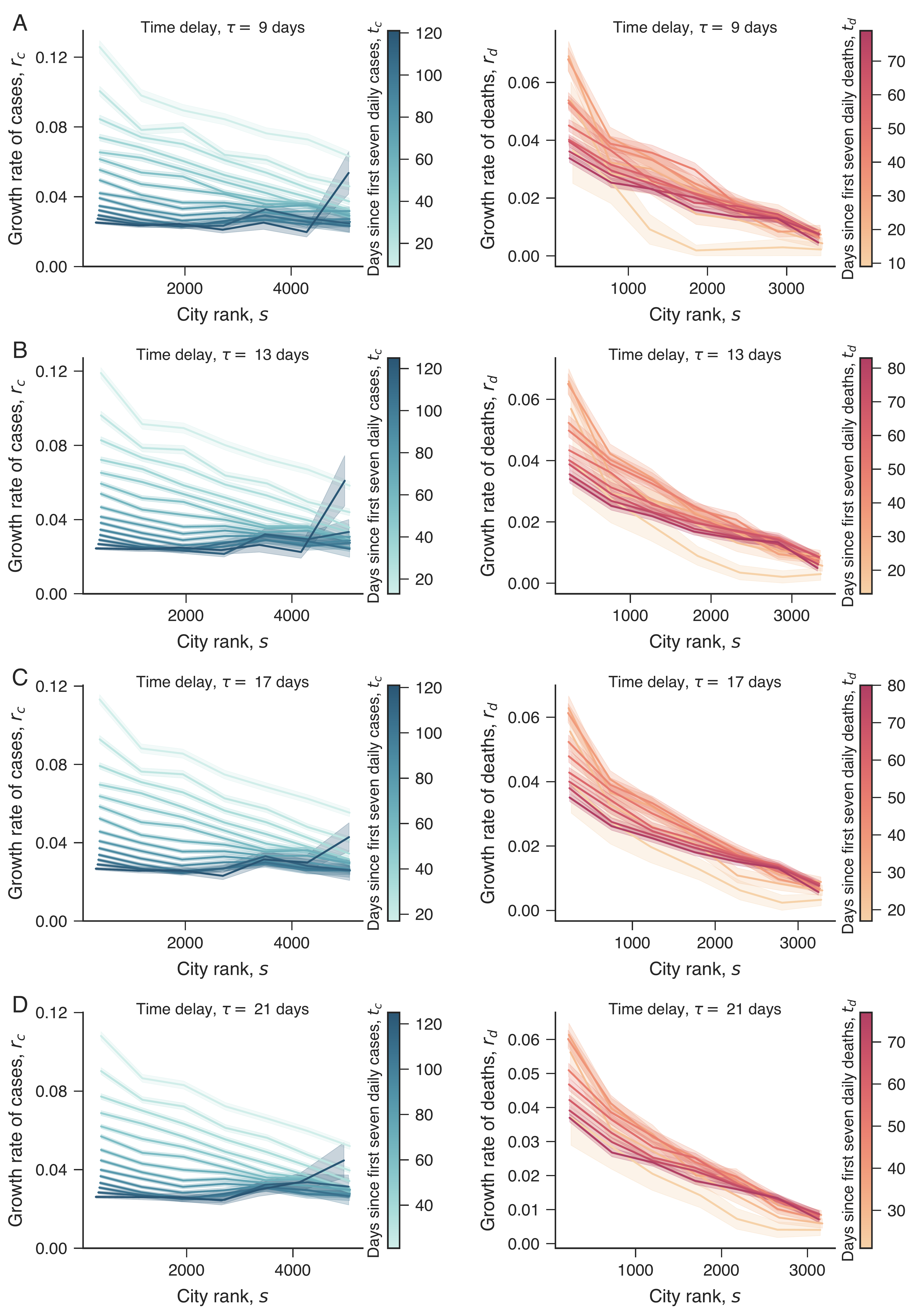}
    \caption{\textbf{Variations in the association between the growth rates and the city rank under different values of time delay $\tau$.} Panels (A)-(D) show the average relationship between the growth rates of COVID-19 cases ($r_c$, left panels) and deaths ($r_d$, right panels) and the city rank $s$ for number of days since the first seven daily cases ($t_c$) or deaths ($t_d$) and for different values of $\tau$ (indicated within the plots).}
    \label{sfig:rank_rates_7}
\end{figure*}

\end{document}